\pdfoutput=1
\documentclass[iop]{emulateapj}

\usepackage{natbib}

\shorttitle{Simulation of Optical Survey Telescopes}
\shortauthors{Peterson et al.}

\begin{document}

\title{Simulation of Astronomical Images from Optical Survey Telescopes using
  a Comprehensive Photon Monte Carlo Approach}

\author{J.~R.~Peterson$^1$, J.~G.~Jernigan$^2$, S.~M.~Kahn$^3$,  A.~P.~Rasmussen$^3$, E.~Peng$^1$, Z.~Ahmad$^1$, J.~Bankert$^1$, C.~Chang$^3$, C.~Claver$^4$, D.~K.~Gilmore$^3$, E.~Grace$^1$, M.~Hannel$^1$, M.~Hodge$^1$, S.~Lorenz$^1$, A.~Lupu$^1$, A.~Meert$^1$, S.~Nagarajan$^1$, N.~Todd$^1$, A.~Winans$^1$, M.~Young$^1$}

\affil{$^1$ Department of Physics and Astronomy, Purdue University, West Lafayette, IN 47907}

\affil{$^2$ Space Sciences Laboratory, University of California, Berkeley, CA
  94720}

\affil{$^3$ Kavli Institute for Particle Astrophysics and Cosmology,
  Stanford University, Stanford, CA 94305}

\affil{$^4$ National Optical Astronomy Observatory, Tucson, AZ 85719}

\email{peters11@purdue.edu}

\begin{abstract}

We present a comprehensive methodology for the simulation of astronomical images
from optical survey telescopes.  We use a photon Monte Carlo approach to
construct images by sampling photons from models of astronomical source populations,
and then simulating those photons  through the system as they interact with the atmosphere, telescope, and
camera.  We demonstrate that all physical effects for optical light that determine the shapes, locations, and brightnesses of individual stars and galaxies can be accurately represented in this formalism.  By using large scale grid computing, modern processors, and an efficient implementation that can produce 400,000 photons/second, we demonstrate that even very large optical surveys can be now be simulated.   We demonstrate that we are able to:  1) construct kilometer scale phase screens necessary for wide-field telescopes, 2) reproduce atmospheric point-spread-function moments using a fast novel hybrid geometric/Fourier technique for non-diffraction
limited telescopes, 3) accurately reproduce the expected spot diagrams for complex aspheric optical designs, and 4) recover system effective area predicted from analytic photometry integrals.  This new code, the photon simulator ({\it PhoSim}), is publicly available.  We have implemented the Large Synoptic Survey Telescope (LSST) design, and it can be extended to other telescopes.  We expect that because of the comprehensive physics implemented in PhoSim, it will be used by the community to plan future observations, interpret detailed existing observations, and quantify systematics related to various astronomical measurements.  Future development and validation by comparisons with real data
will continue to improve the fidelity and usability of the code.

\end{abstract}


\section{Introduction}

\subsection{Astrophysical Instrument Simulations}

Modern survey astronomy is leading to extremely large samples of source populations amenable to subtle statistical analyses.  The precision astrophysical investigations that are thereby enabled require high fidelity simulations to properly
interpret the observations.  In addition to detailed simulations of astrophysical environments (e.g. \citealt{fryxell2000};
 \citealt{springel2005}; \citealt{stone2008}),
it is increasingly important to build high fidelity end-to-end
simulations of the instrumentation as well.  This can be important for design and optimization of the
telescope and camera, for planning future astronomical observations, and for building
sophisticated analysis software prior to data taking.  In addition,
simulated data can be processed along with the actual observed data through the same analysis pipelines, so that
the flaws, biases, and efficiencies of the analyses can be determined accurately.   Such an approach is now an
indispensable part of modern particle physics (\citealt{agostinelli2003};
\citealt{allison2006}), high energy astrophysics (\citealt{peterson2004}; \citealt{peterson2007};
\citealt{andersson2007}; \citealt{davis2012}; \citealt{ackermann2012}), optical astronomy using adaptive optics (\citealt{lane}, \citealt{ellerbroek}, \citealt{lelouarn}, \citealt{britton}, \citealt{jolissaint}), and has recently become more common in optical survey astronomy (\citealt{bertin},
\citealt{dobke}, \citealt{mandelbaum}).   In this work, we outline a method
for high fidelity optical astronomical image simulation appropriate for survey telescopes.

\subsection{Goals for the Simulation}

The goal of this work is to produce high fidelity simulated optical astronomical images.  The level of detail necessary to achieve high fidelity can be precisely defined by considering the measurable properties of the images that we are interested in reproducing.
We can divide those measurable image properties into two categories:  1) primary image properties and 2) secondary image properties.
The primary image properties are:  the point-spread-function (PSF) full width at half maximum (FWHM), the photometric zeropoint,
the plate scale (or astrometric scale), and the background intensity level.  These quantities are predicted from physical effects in the instrument/atmosphere system.  If we are only interested in simulated images that correctly reproduce those four primary properties,
it is possible to use an empirical parametric approach.  For example, the photometric zeropoint and the object's flux within a
band can be used to calculate an object's intensity.  The PSF FWHM and the object's spatial distribution yield its observed morphology.
The astrometric scale determines the location in the image, and the background provides the extra uniform intensity level.
In many cases, such a simulator would be sufficient to predict what a basic image would look like from an astronomical telescope.  However, a simplified simulation tool that only matches these four properties is insufficient for detailed applications.

It is significantly more difficult to predict more complex secondary image properties
using parametric models.  Examples include:  the PSF radial profile, the PSF size wavelength dependence, the spatial
variation of the PSF size, the PSF shape (notably,
ellipticity), the PSF shape wavelength dependence, the PSF shape spatial
decorrelation, the PSF shape spatial variation, the differential astrometric
shift non-linearity with field angle, the differential astrometric
non-linearity with wavelength, the differential astrometric decorrelation with
angle, the differential astrometric decorrelation variation, the photometric
chromaticity, the photometric variation in time, the photometric variation with field
angle, the background variation in time, the background spatial dependence,
and the background wavelength dependence. Understanding such properties is essential for subtle statistical analyses
associated with galaxy shape measurements, stellar astrometry, and precision photometry. Some of the astrophysical investigations
that require such analyses include:  photometry of supernovae for measuring the
expansion rate of the Universe (\citealt{riess1998}, \citealt{perlmutter1999}),
using stellar proper motions and parallax measurements to map the structure and kinematics of the Milky Way
(\citealt{hoeg2000}; \citealt{monet2003}), measurement of large
scale galaxy structure statistics (\citealt{eisenstein2005}),
and studies of weak gravitational
lensing for both dark matter and dark energy investigations (\citealt{tyson1990}).

In this paper, we describe a simulation tool that takes explicit account of all the
detailed atmosphere, telescope, and camera physical effects that determine the
propagation of light to the focal plane.    We show that the relevant physics can be most efficiently encoded in terms of photon
manipulations.
The goal of this work is to be able to predict all of the primary and
secondary image properties from physical models of the atmosphere and
instrument within the context of a photon Monte Carlo approach.
We use very few approximations, and the overall simulation accuracy
has been carefully controlled to ensure that calculational errors are small in comparison to
the statistical uncertainties associated with measurements
of typical sources in the survey.

\subsection{Simulation Regime for Optical Surveys}

In this work, we are interested in simulating images containing large numbers of astronomical objects (typically, millions) in the visible and near infrared wave-band ($\sim$300 to $\sim$1200 nm) that would be obtained in an optical survey.  A complete physical simulation with no approximations would take account of the detailed material
properties of every element of the telescope and camera and their response to
all external forces and temperature variations, a complete hydrodynamic treatment of the atmosphere, and
simulation of the full quantum mechanical nature of light.
However, such a rigorous approach is unwieldy, and unnecessary, since a number of key simplifying approximations can be
introduced for optical survey telescopes which are non-diffraction limited.  A telescope is non-diffraction limited if

$$\mbox{max} \left[ \frac{\lambda}{r_0}, \frac{\alpha}{F} \right] >> \frac{\lambda}{D}$$

\noindent
where $\lambda$ is the wavelength, $D$ is the diameter of the
telescope, $F$ is the focal length of the telescope, $r_0$ is the Fried
parameter, and $\alpha$ is the characteristic size of aberrations introduced by
the telescope and camera.  The
Fried parameter describes the characteristic beam diameter where the optical phase rms value is close to 1 radian (\citealt{fried1965}).  Thus, the first ratio
is associated with distortions introduced by optical turbulence (usually named seeing), while the second accounts for imperfections in
the optical system.

It is convenient to
rewrite this condition in the form:

$$\mbox{max} \left[ \frac{D}{r_0}, \frac{\alpha}{\lambda~f} \right]  >> 1$$

\noindent
where $f$ is the focal ratio of the system, $F \// D$.
For most ground-based optical telescopes the value of $r_0$ lies between 5 and 30 cm,
depending on atmospheric conditions.  For modern survey telescopes like LSST (8.4 m diameter primary), the first
ratio is much larger than unity\footnote{ The Fried parameter depends on wavelength as $\lambda^{\frac{6}{5}}$, so throughout this work, values are referenced to 500 nm as this is the standard convention in the literature.}.  In addition, to achieve wide fields of view over a range of wavelengths, practical issues cause these telescopes to be
somewhat aberrated, so that the second ratio is generally greater than unity as well.
If some of the physical effects (e.g. either large aberrations from atmospheric turbulence or large telescope optics aberrations) cause the system to be non diffraction-limited, we can consider standard ray optics techniques for those parts of the photon simulation.  This does not preclude the use of quantum mechanical calculation or wave-like interference computational techniques at the appropriate physical places as we discuss in \S2.1.

\subsection{LSST}

The initial implementation of the data describing a telescope and site in PhoSim is LSST.
LSST is a wide-field ground-based telescope with extremely large
etendue (the product of the effective area and the field of view), 320
$\mbox{m}^2 \mbox{deg}^2$, roughly a factor ten higher than all previous facilities.  The large etendue will enable
an unprecedented survey of the optical sky.  It also results in a very high anticipated data
rate of 15 Terabytes of images per night.  LSST is an extreme case
of the optical survey regime discussed above.  It is also the most
challenging telescope to simulate computationally because of the large quantity of data.  We have
therefore chosen to construct the simulator to simulate LSST and
focus on that application primarily in this paper.  However, we have written the
simulation code to keep the LSST design data separate from the physics
algorithms, so other telescopes can be simulated with this code as well.

LSST will be constructed on the El Pe\~{n}\'{o}n peak at the Cerro Pach\'{o}n ridge in Chile at
2660 m above sea level.  The site is expected to deliver a median seeing of 0.67
arcseconds (with 0.44'' as the 25th percentile and 0.81'' as the 75th percentile).
LSST is an 8.4 m optical telescope with a three mirror modified Paul Baker design to allow for a large unaberrated field of view.  It
has an effective f/\# of 1.23.  It has three correcting lenses and complement of
filters as part of a 3.2 gigapixel camera.  The camera has 189 individual 4k x 4k CCDs.  The 6 filters (u,g,r,i,z,y) cover the
wavelength region from 300 to 1100 nm.   The intrinsic aberrations of
the optical system, the residual aberrations after compensation of the active optics control system using curvature sensors,
and the charge diffusion in the devices is expected to contribute about 0.4
arcseconds to the PSF (\citealt{ivezic2008}; \citealt{lsstreference}).

\subsection{Scope and Interfaces}

The simulation work described here is embodied in a publically available code called the Photon Simulator or PhoSim.  The code and the documentation are located at https://www.bitbucket.org/phosim/phosim\_release.  PhoSim is a stand-alone code that requires catalogs of astrophysical objects (positions, fluxes, shape, spectra, etc.) and operational parameters (pointing information, instrument configuration, etc.) as input.  The output of PhoSim is a stream of images.  The user can create the inputs to either match a real observation or to create a hypothetical observation.  The combination of the operational parameters and astrophysical catalog data is called an {\it instance catalog} and is described in \S3.  Internally, PhoSim uses the physics of the atmosphere, telescope, and camera to predict high fidelity images based on the input and is described in the following section.

\section{Physics Description and Algorithms}

In the following, we describe the complete set of atmosphere/instrument physics that is encapsulated in the PhoSim code.  We describe how the physical interactions can be computed in terms of photons manipulations.

\subsection{Basic Monte Carlo Methodology}

Given the simulation regime (\S1.3) and the simulation goals (\S1.2) it is
ideal to adopt a Monte Carlo approach.  The basic Monte Carlo approach
has been an essential part of computational physics for some time
(\citealt{ulam}; \citealt{metropolis}; \citealt{ulam2}).  In our case, the
Monte Carlo is implemented by following individual astrophysical photons.  Thus the
integral over the time-dependent, angle-dependent, and energy-dependent incident
radiation field is accomplished by simply following individual photons throughout the
atmosphere/instrument system.  This is particularly important because a large
fraction of the physical effects have wavelength and angle-dependent
properties.

In certain contexts, the photon's propagation can be represented using ray
optics.  However, as we show in this formalism, this does not preclude us from using a diffraction calculation
for perturbing elements with a wavefront shift smaller than the photon wavelength, a quantum mechanical
calculation for photon-atom interactions, or an electromagnetic wave
interference calculation at interface boundaries.  Thus, ray optics is the
methodology we use to represent the photon's interaction with physical elements where ray
optics is appropriate, whereas more complex wave optics (or quantum mechanical
physics) can describe the photon in other areas.

In ray optics, the photon state can be represented by a set of
eight numbers (ignoring polarization\footnote{Polarization is relevant in interference calculations that result in fringing.}):   a vector describing the photon's current position ($\vec{x}$), a
unit-vector describing the angle of propagation ($\hat{n}$), a time stamp when the photon
arrived at Earth ($t$), and a wavelength for the photon ($\lambda$).  Throughout the simulation
of the physics, 3 possible alterations of the photon's variables may occur:  1) the photon's
trajectory may change and therefore the unit-vector changes its value, 2) the
photon's position may change as it propagates along its path, and 3) the
photon may be removed (most likely because it scattered in an uninteresting
direction).  To a good approximation, the photon's wavelength does not change
and the time stamp of the photon can be regarded a constant, since a photon
moves from the top of the atmosphere to the detector in less than 30 $\mu s$
and none of the physical models have rapid changes on that time-scale.
By the end of the calculation, we may have converted the photon into an
electron in the detector.  The electron similarly can be described by the same
formalism, except that the magnitude of the velocity is the relevant quantity
instead of the wavelength.  We are therefore using a photon Monte Carlo
method to simultaneously manage the change in trajectory as well as the
change in intensity or throughput.

\subsection{Sky Photon Sampling}

The first step of the simulation is to create the photons from astrophysical
sources.  This involves populating the quantities describing a photon.  The
propagation direction for a photon coming from a source at a position
($\alpha_i, \delta_i$) is determined by a unit vector,
$\hat{x}, \hat{y},$ and $\hat{z}$ that is calculated from the current bore-sight of the telescope ($\alpha, \delta$) according to

$$ \hat{z} = (\cos{\alpha} \cos{\delta}, \sin{\alpha} \cos{\delta}, \sin{\delta}) $$

$$\hat{x} = (\cos{\left(\alpha+\pi/2\right)}, \sin{\left( \alpha+\pi/2 \right)} ,0) $$

$$ \hat{y} = \hat{z} \times \hat{x}$$

\noindent
where $\times$ represents a vector cross product.  The coordinate system is then in the telescope's frame where $\hat{z}$ is the optical axis.  We calculate the angle of propagation by first calculating a source vector, $\hat{s}$

$$ \hat{s} = (\cos{\alpha_i} \cos{\delta_i}, \sin{\alpha_i} \cos{\delta_i},
\sin{\delta_i}) $$

\noindent
and convert the vector into tangent coordinates by

$$ \hat{n}_x = \frac{ \hat{s} \cdot \hat{x} }{ \hat{s} \cdot \hat{z}}~~;~~\hat{n}_y = \frac{ \hat{s} \cdot \hat{y} }{ \hat{s} \cdot \hat{z}}~~;~~\hat{n}_z = -\sqrt{1-\hat{n}_x^2-\hat{n}_y^2} $$

\noindent
where $\cdot$ indicates a dot product.  The source's position is assumed to have astronomical aberration included.
The time stamp, $t$ of the photon is chosen in a Poisson manner by considering the total exposure
time, $t_e$ and a uniform random number, $u$

$$t = t_0 + t_e u$$

\noindent
where we include a time offset, $t_0$, if we are simulating a sequence of
exposures.  This will affect other quantities in the case of transient or moving objects.

We determine the photon's initial position, $\vec{x}$, by first sampling
the annular pupil of the primary mirror by choosing the photon's position in polar
coordinates,

$$ r=\sqrt{ u \left( r_{o}^2 - r_{i}^2 \right) +
  r_{i}^2 }~~;~~\phi = 2 \pi v$$

\noindent
where u and v are uniform random numbers, $r_{o}$ is the outer radius and
$r_{i}$ is the inner radius.  The outer and inner radius can be chosen to be slightly larger than
 the actual physical aperture to allow for photons that scatter into the aperture.
Then, we calculate $x=r \cos{\phi}$ and
$y=r \sin{\phi}$, and determine the $z$ position by using the mirror surface
function, $z(x,y)$.  Then, we move the photon to the top of the atmosphere
$\vec{x} \rightarrow \vec{x} + \hat{n} l$.   $l$ is calculated using
$-\frac{h_0-z}{\hat{n}_z}$ where $h_0$ is the top of the atmosphere (100 km).

We choose the wavelength of the photon by sampling from a spectral
energy distribution describing the source.  Spectral energy distributions (SEDs) are
normally quoted in terms of flux units (ergs cm$^{-2}$ s$^{-1}$ $\mbox{\AA}^{-1}$) at given
wavelength interval in the spectrum.  We can convert this into a relative probability of a
photon appearing in a given wavelength interval by dividing by the average photon
energy and multiplying by the wavelength interval.  Then we sample from this
distribution by converting into a cumulative distribution and drawing a
uniform random number.  We blur the wavelength by the bin width.  By convention, we
use the part of the SED with wavelength less than 1200 nm (which would be appropriate for LSST).  We also draw the
photon in the rest frame of the object, and then redshift the photon by
multiplying the wavelength by 1+z.  This allows us to re-use the SEDs for different
galaxies at different redshifts, and still obtain a diversity of galaxy colors.

\subsubsection{Non-Point Source Models}

Sources that are extended on the sky are treated by first choosing the
direction, $\hat{n}$, using the procedure above for the nominal center
($\alpha_i$, $\delta_i$) of the source emission, but then perturbing the unit
vector by a structural model for the spatial emission.  We can use a paraxial ray approximation to perturb the values of $\hat{n}_x$ and $\hat{n}_y$ and
then renormalize the unit vector.

The most useful spatial model we have constructed is an ellipsoidal Sersic
distribution.  The intensity of light of the Sersic distribution (\citealt{sersic1963}) is given by

$$I(r) = I_0 e^{-b_n \left( \frac{r}{r_0} \right)^{\frac{1}{n}}} $$

where $r_0$ is the scale radius, $n$ is the Sersic index, and $b_n$ is a
normalization constant for each $n$.  We then draw a
value of $r$ from this probability distribution, and then choose the distance
along the major and minor axes as $ar \// r_0$ and $br \// r_0$ where
$a$ and $b$ are model inputs.  Finally, we rotate the major and minor axes by
a rotation angle, $\phi$.  In practice, we have found it useful to represent galaxies as a pair of
ellipsoidal Sersic models where one represents the bulge and one represents
the disk.  Each component has a different Sersic index, their centers are
possibly offset, and they can have different major and minor axes.  A pair of ellipsoidal Sersic models results in a 12 parameter model with reasonable fidelity.

For more accurate galaxy morphology simulations, another common model we
include is to simply input {\it truth} images of the source
before it would have been observed through the effects of the atmosphere and
telescope.  We use this by using this truth image as the 2 dimensional probability
distribution of finding a photon in a given angular pixel.  We then select photons from
this distribution.  We also allow for an arbitrary rotation and scaling that places the
source at different distances and orientations.  The truth images can either come from other telescopes
(particularly those taken with a night with better seeing or a space-based telescopes), or be generated by another simulation.

A simple perturbation to the spatial models is to include the effect of weak
gravitational lensing for distant galaxies.  This is accomplished by taking
the photon's relative position to the source's center $(\delta
x, \delta y)$ and apply the matrix

$$ \left( \begin{array}{cc} 1-\gamma_1-\kappa & \gamma_2 \\ \gamma_2 & 1+\gamma_1-\kappa
  \\ \end{array} \right) $$

\noindent
where $\gamma_1$, $\gamma_2$, and $\kappa$ are the usual weak lensing shear and convergence parameters.  The user's cosmological simulation can estimate these parameters from the individual foreground dark matter distribution for each galaxy individually.
We handle moving objects (e.g. Solar System objects or satellites) by simply
perturbing the initial photon direction by a proper motion vector (in units
of angle per time) and multiplying by the photon's relative time stamp.  This results in streaks when the full simulation is completed.  Both of these effects can also be handled by a distorted truth image, but are added options since they can be applied to objects when expressed as catalog entries.

\subsubsection{Diffuse Emission Simulation}

There are a large number of photons that contribute to the optical sky
background.  We model these photons from 3 sources:  airglow from the dark
sky, reflected moonlight, and additional emission near twilight.  We also
simulate light from a dome screen through the same mechanism.  For each of
these forms of diffuse light, we model them by a collection of sources
uniformly spaced on the celestial sphere.  This is necessary to allow for complex spatial gradients.  We space the sources by 15
arcseconds, and draw photons from a two dimensional spatial gaussian having a standard deviation of
15 arcseconds\footnote{For arbitrarily high statistics this approach will have a periodic wave artifact, however, for practical background levels the spacing of the sources is sufficient to make this unobservable.}.  This results in a statistically uniform illumination pattern, but limits us
to not have a spatial variation of the illumination patterns on scales smaller than 15
arcseconds.  For each diffuse model, we therefore need to predict the
intensity of light across the field on 15 arcseconds scale.

For the simulation of airglow emission, we use a spectral energy distribution
taken from \cite{patat2006}.  The overall intensity for a given exposure is simulated
as having a r band magnitude from a gaussian distribution with mean of 22.08
and a standard deviation of 0.9.  We then predict a relative variation across the field
using spatial power spectrum measurements of \cite{adams} where the variation
was proportional to $k^{-0.3}$ and normalized to be 3\% at 1 degree.  We also increase
  the emission in proportion to the zenith angle (\citealt{adams}).
The mean sky background and its variation are consistent with the
measurements of \cite{krisciunas1991}.

The moon's intrinsic brightness as a function of its phase and altitude
follows the calculation of \cite{krisciunas1991}.  We use an empirical lunar spectrum for its SED.  We then need to predict the
brightness where the telescope is pointing.  Here we use the \cite{krisciunas1991} formula that has terms for the Rayleigh and Mie scattering of the moonlight.  \cite{krisciunas1991}
only calculated the lunar brightness for one band so we simply scale the
Rayleigh term by inverse wavelength to the fourth power.  Conversely, Mie scattering is approximately
wavelength independent.  The sky brightness is increased near twilight according to the sun's altitude
using a color-dependent model of \cite{patat2006}.

\subsubsection{Number of Photons}

The total number
of photons for a particular source is calculated by considering the AB magnitude at a particular
wavelength.  We have found it most convenient to normalize the spectrum at a
wavelength of 500 nm divided by $1+z$ of the source.  We then convert
the flux nearest that wavelength to units of ergs cm$^{-2}$ s$^{-1}$ Hz$^{-1}$.
Then, we convert the SED to a relative fraction of
photons in each bin.  Using the probability of finding a photon in the bin near the
reference wavelength, 500 nm/$(1+z)$, one can then calculate the total
number of photons per sq. centimeter per second from that source at all wavelengths.  This convention is most useful using a un-redshifted spectrum, since the same SED can be used for multiple sources at different redshifts.  Conversely, this does not preclude a user redshifting a SED before input and then setting the redshift to zero.  The total number of photons (without including any efficiency losses) is calculated by multiplying by the aperture
area and exposure time.  We then modulate that expected photon count rate using a Poisson distribution.

\subsubsection{Dust Models}

A final step of the sky simulation is to remove some fraction of the photons
due to dust absorption.  We use two dust extinction curves: the \cite{ccm1989} model and
the \cite{calzetti2000} model.  Each model has two parameters:  $A_V$ and
$R_v$.  The models are stored in a variety of grid lookup tables and the dust
is applied by first calculating the optical depth, $\tau$, using the photon's
wavelength.  Then we destroy photons with probability $e^{-\tau}$.
Performing dust extinction through a Monte Carlo approach conveniently also avoids
construction of a unique SED for every single source.  For extra-galactic
sources, we have the option of applying the dust absorption both before and
after the redshift of the photon, representing absorption in the galaxy as
well as in the Milky Way.

\subsection{Atmosphere Simulation}

After creation of the photons from the astrophysical and sky emission, we
then propagate the photons through an atmosphere simulation.  We first describe the
structure of the atmospheric components, and then describe how the photon
interacts with those components.

\subsubsection{Atmosphere Structure}

We model the atmospheric structure by a series of plane-parallel layers.  Each
layer covers the absorption and turbulence blurring between two altitudes.
The propagation of a ray from one layer to the next is straightforward using plane-parallel layers.  The photon's position can be
updated using $\vec{x} \rightarrow \vec{x} + \hat{n} l$ where the scalar
distance, $l$ is calculated from

$$ l = \frac{h_i - h_{i+1}}{-\hat{n}_z}$$

\noindent
where $h_i$ is the altitude of the interface between the two layers and $\hat{n}_z$ is the same as in \S2.2.

Atmospheric turbulence is known to significantly degrade overall image quality
of astronomical telescopes.   The energy power spectrum of atmospheric turbulence is
known to have a power spectrum that approximately follows the scale-free Kolmogorov
spectrum ($k^{-11/3}$), so the largest powers are on the largest
scales (\citealt{kolmogorov}).  This spectrum extends down to the viscous limit
(a few mm) and up to the scale where the turbulence is driven.   The complete spectrum is often parameterized by an outer scale, $L_0$, and a flat power spectrum above that point (\citealt{vonkarman}),

$$\left( k^2 + L_0^{-2} \right)^{-11/6}$$

\noindent
The justification of using this spectrum to describe the real atmospheric turbulence has
been established in detail (\citealt{tokovinin2007}).  The use of a von Karman
spectrum rather than a Kolmogorov spectrum has several observational consequences (\citealt{martinez}).
Other more complex power spectra are sometimes considered (\citealt{hill}) and
non-Kolmogorov effects are an active area of turbulence research,
but astronomical image quality is generally not particularly
sensitive to these details.

The temperature variation produced by this airflow turbulence pattern leads to
index of refraction variations that affect light propagation (see e.g. \citealt{roddier}).  The
deviation of index of refraction from unity ($n-1$) is approximately linearly
proportional to the pressure and inversely to the temperature, and the effect of pressure fluctuation is negligible compared to the temperature variation.  So therefore the same turbulent eddies produce index of refraction variations with the same spectrum as the temperature variation. The entire effect of the
turbulence on light propagation is then represented by a phase
shift as a function of spatial position (i.e., a phase screen).

Furthermore, the turbulence pattern itself for a particular layer is known to not change significantly during the time it takes to cross the aperture of the telescope (a fraction of a second).  To a good approximation, the pattern tends to drift with some wind velocity vector and remain essentially frozen during its aperture crossing
(Taylor hypothesis; \citealt{taylor}, \citealt{favre}, \citealt{poyneer}).  In addition, the most significant turbulence tends to
occur stochastically in relatively narrow interfaces due to either
differential shearing of the atmosphere or vertical instabilities.  Hence, it is very common to
represent the turbulence as a series of {\it frozen} plane-parallel
two-dimensional screens where the three-dimensional structure for a given slab
of the atmosphere has already been collapsed into two dimensions.

We therefore propagate light through a series frozen-phase screens
drifted by wind velocity vectors.  This simulation approach is standard in the
adaptive optics community and is well studied (e.g. \citealt{lane}, \citealt{ellerbroek}, \citealt{lelouarn}, \citealt{britton}, \citealt{jolissaint}).  However, we made two novel innovations to suit our
particular problem that are not standard techniques.
 First, instead of propagating light through by computing the full
diffraction integral, we used a novel geometric raytracing approximation for
the low-frequency part of the phase screen described
in \S2.3.2.
Second, we constructed the phase screens on four different
scales and repeated the three smaller phase screens on the larger
scales.  The tiling scheme has been previously explored by
\cite{vorontsov} for a single scale, and here we simply extend the method to
four scales.  This
general approach is necessary to capture the large scales involved in the
simulation of a wide-field telescope like LSST.  In particular, the screen
size, $L$ has to be large enough to not repeat during an exposure time
($L>v_{\mbox{wind}} t_{\mbox{exp}}$) and large enough to not repeat when
off-axis sources are considered ($L>h \theta$, where $h$ is the height of the
highest layer). We therefore use a linear superposition of four 1024 by 1024 pixel screens
having a pixel size of $1$, $8$, $64$ and $512$ cm.  Every pixel in the 5 km x
5 km screen is therefore unique.  Numerical artifacts result from certain alignments of
patterns along the fundamental axes of the screens with the wind direction (0,
$\frac{\pi}{2}$, $\frac{\pi}{4}$, $\arctan{\frac{1}{2}}$, etc.) since the
patterns may be repeat and a small part of the screen gets used.  However, the
actual atmosphere has about a 5 degree per minute drift of
the wind direction (\citealt{mahrt}), which makes these artifacts not occur in
the complete simulation when this effect is turned on.  This is particularly important in the detailed simulations of
wavefront images.

We drift the turbulence screens according to a wind model where we predict a
wind vector as a function of height for a particular observation.
In order to determine where a photon hit a screen at a given layer, we first
calculate the x and y position.    We then calculate the pixel in the
appropriate screen given two components of the wind vector for each screen.
The arrival time of the photon then dictates exactly which pixel is used.
Using the NOAA NCEP/NCAR Reanalysis Monthly Database for a particular location, we fit the historical data to a Weibull
distribution (k=2) for the wind velocity.  Both the wind direction and
magnitude have a seasonal distribution.  We drift the wind
direction during an exposure according to measurements of \cite{mahrt}.

In order to construct the turbulence screens, we require the single parameter,
the outer scale, of the \cite{vonkarman} model for each screen.  Within the
context of this model, it is often argued that this varies as a function of
height, has a large time variation, and likely has a log-normal distribution
(e.g. \citealt{beland1988}, \citealt{coulman1988}, \citealt{abahamid2004}).  A log-normal distribution is a good match to the LSST site where the mean (35.6 m) and
median (26.7 m) parameters were measured by
\cite{boccas2004}.   The possible altitude-dependence of the outer scale is still an active area of research (e.g \citealt{abahamid2004}), so we do not have an altitude-dependence to the outer scale.

To predict the relative turbulence intensity as a function of altitude, we construct an interface based on the model of \cite{tokovinin2006} for the Cerro Pachon site.  However, other sites can clearly be represented by different set of parameters as is evident in the general nature of the parameterization below.
In this model, they use a 7 layer model (a
ground layer and 6 layers at 0.5, 1, 2, 4, 8, and 16 km).  With a series of
measurement they then quantified the 25th, 50th, and 75th percentile of the
turbulence integral, $J=\int C^2_N (h) dh$, where $C^2_N$ is the refractive index structure function and $h$ is
the altitude.  We then use these percentiles and represent the complete
distribution of $J$ at each altitude as a lognormal distribution.
\cite{tokovinin2006} found that the turbulence intensity of the ground layer
was largely independent of the other layers.  The free atmosphere layers,
however, were highly correlated as seen by comparing Figure 5 and Table 3 in
\cite{tokovinin2006}.  Therefore, we use a single random number for the free
atmosphere and a single random number for the ground layer.  This
then tends to produce poor seeing when either the free atmosphere or the ground
layer is particularly turbulent.  We can use this model in two different
ways.  First, if a random atmosphere is desired, we simply draw turbulence
intensities from this model.  Second, to obtain a particular seeing, we run
this model hundreds of time to produce the desired seeing.  In either case,
the final turbulence intensity values are used to normalize each screen.

In addition to the screens of turbulence, we track four components that represent
the bulk air located between the screens: the density of the atmosphere
vs. height, the density of water vapor vs. height, the density of molecular
oxygen vs. height, and the density of ozone vs. height (\citealt{sander2006}; \citealt{thomas1999}).  These profiles are then used to calculate the column
density between layers (\citealt{bodhane1999}, \citealt{liou2002})
 We calculate the column density relative to the appropriate altitude.  We also modulate the overall column density from one simulation to
 the next by a lognormal distribution with width equal to 0.18, 0.20, 0.002,
 and 0.01 for water, ozone, molecular oxygen, and the overall density,
 respectively.  This represents the natural time-dependent variation.
When the telescope is not pointed at zenith, we increase the column density by the exact zenith-dependent airmass.  We recalculate this for every photon
separately since the variation of airmass across the field can
be significant for wide fields.  We then use these column densities to
calculate the optical depths, which is described in \S2.4.2.  The local column density is also perturbed by a factor of
$p \times \sqrt{N_{\mbox layers}}$ of a screen with an arbitrary variation pattern at
each layer.  In this way, the opacity will vary slightly from exposure
to the next, and it will vary across the field in a complex way with
amplitude, $p$.  However, the parameterization of reasonable variation patterns is still an active area of research.  We also have clouds screens to represent the relative opacity of cloud
absorption at various heights.  The cloud screens have an exponential
structure function with a angular coherence scale of 2 degrees to be
consistent with the measurements of \cite{ivezic2007} based on SDSS wide
field images.  We assumed that the structure is isotropic, but clearly have non-isotropic
structure that could be related to the wind direction.  We use two layers of
clouds each with a different wind velocity to create a partly realistic complexity that
would at least mimic the difficulties for photometric calibration.  The
photon interactions with all the components and properties of the atmosphere are described in the following sections.

\subsubsection{Photon Interactions with the Turbulent Atmosphere}

An important component of the atmosphere simulation is the propagation of the
photons through the turbulent screens.
The theory of how light propagates through a turbulent medium has been
extensively studied (\citealt{tatarski}, \citealt{fried1965},
\citealt{clifford}, \citealt{roddier}, \citealt{schmidt}), and it is common to use
 Hugyens-Fresnel scalar diffraction theory,  which represents light
as a monochromatic wave having a scalar amplitude at all spatial positions.
The so-called {\it geometric optics limit}, where non-linear effects can be ignored, is valid when the characteristic height of turbulent layers in the atmosphere is less than the square of the characteristic turbulent cell size divided by the wavelength in which the observation is performed

$$ h \leq \frac{l^2}{\lambda} $$

\noindent
For typical atmospheric conditions at good sites, h $\approx$ 10 km, and l $\approx$ 10 cm, which means that this condition is essentially satisfied across the entire optical band:  $\lambda$ = 0.3 to 1 $\mu$m.    In this limit, we can write the PSF in terms of the Fraunhofer integral (see e.g. \citealt{goodman}).  Here, the instantaneous PSF is given by the square of the Fourier transform of the electric wave amplitude\footnote{The electric wave amplitude, $U(\vec{r})$, can be written as $U(\vec{r}) = E(\vec{r}) e^{i \phi(\vec{r})} e^{i \omega t}$.  However, for the exposures considered here the amplitude fluctuation is negligible and the last factor has no effect on the image, so only the second factor is relevant in the above integral.} over the pupil plane

$$ PSF(\hat{n}) = {\vline \frac{1}{A} \int_A d^2\vec{r} e^{i k \hat{n} \cdot \vec{r}} e^{i \phi(\vec{r})} \vline}^2$$

\noindent
where k = $2 \pi / \lambda$, $\hat{n}$ is the unit vector pointing to that particular point on the focal plane, and $A$ is the telescope pupil transmission.  For the usual small angular displacements appropriate to the PSF, $\hat{n}$ has coordinates ($\theta_x$, $\theta_y$, 1), where $\theta_x$ and $\theta_y$ are coordinates in the focal plane with units of angle.  The phase, $\phi$, which appears in this expression is the total phase shift produced by all of the atmospheric layers along the line of sight due to variations in the index of refraction.  Note that the PSF can also be expressed as the Fourier Transform of the correlation function of the exponential of the complex phase

$$ B ( \vec{\rho} ) \equiv \frac{1}{A} \int_A d^2\vec{r} e^{i \phi(\vec{r})} e^{-i \phi(\vec{r}+\vec{\rho})}$$

\noindent
The turbulence in the atmosphere is a stochastic process, which means that both the phase correlation function and the PSF will depend strongly on time through the variation in the random structures of the various atmospheric layers that cross the pupil plane at any particular instant.  However, for an exposure of finite duration, a larger footprint on these atmospheric layers is observed, which means that the effective PSF is averaged over such structures, and it is meaningful to estimate it via statistical techniques.  In the adaptive optics literature, it is common to talk about two distinct limits:  the long exposure limit, in which one takes a complete average over the turbulent structure on all spatial scales, and the short exposure limit, in which one takes some average over scales which are small compared to the diameter of the pupil, but ignores the effects of much larger scales.  The latter is appropriate, because as we show below, the primary effect of the larger scales for short exposures is image displacement, not an increase in PSF width.

If we are interested in estimating PSF anisotropy and decentering for moderate exposures, an intermediate limit is required.  Since there are no preferred axes in the problem, the long exposure limit, which involves a full statistical average over all of the turbulent structures must clearly yield a circularly symmetric PSF with no decenter.  However, there are significant variations in the image motion for moderate exposures, which are in fact the principal causes of residual anisotropies, so these must be modeled correctly.  That means that the short exposure limit is also inappropriate.

The intermediate limit can be found by separating out the contributions to the PSF from structures on large and small scales.  In what follows, we provide a formal justification for how that works in detail.  We first define the separate contributions to $\phi$ that are contributed by large and small scales

$$ \phi (\vec{r}) = \phi_> (\vec{r}) + \phi_< (\vec{r})$$

\noindent
where $\phi_> (\vec{r})$ is the contribution from structures with wavenumber larger than some critical value, $\kappa_{\mbox{crit}}$, and
$\phi_< (\vec{r})$ is the contribution from structures with wavenumber less than $\kappa_{\mbox{crit}}$.   We choose $\kappa_{\mbox{crit}}$ such that structures with larger wavenumbers are very well sampled during the exposure, so that the long exposure limit is indeed applicable on those scales, and we can estimate their contribution statistically.  For wavenumbers less than $\kappa_{\mbox{crit}}$, we effectively compute $\phi_< (\vec{r})$ directly for a simulated exposure by generating phase screens with appropriate resolution, i.e. with a sampling $2 \pi / \kappa_{\mbox{crit}}$.

Note that because the PSF is insensitive to the mean phase, both $\phi_< (\vec{r})$ and $\phi_> (\vec{r})$ can be taken to have mean zero, with no loss of generality.  Then, the correlation function, $B (\vec{\rho})$, that we introduced earlier, can be written in the form

$$ B (\vec{\rho}) = \frac{1}{A} \int_A d^2 \vec{r} e^{i ( \phi_> ( \vec{r} ) - \phi_> ( \vec{r} + \vec{\rho})} e^{i ( \phi_< ( \vec{r} ) - \phi_< ( \vec{r} + \vec{\rho})} $$

\noindent
As indicated, we evaluate the first complex exponential term in the integral by taking a statistical average of turbulent structures at higher wavenumbers.  That clearly removes an explicit dependence on $\vec{r}$, so it is possible to express the correlation function as the product of two separate correlation functions for the two separated wavenumber regimes

$$ B ( \vec{\rho} ) = B_> ( \vec{\rho} ) B_< ( \vec{\rho} ) $$

\noindent
The PSF due to the atmosphere is the Fourier Transform of the correlation function.  It can therefore be represented as the convolution of the separate PSF contributions from large and small wavenumbers

$$ PSF ( \hat{n} ) = \int d^2 \hat{m} PSF_> ( \hat{m} ) PSF_< (\hat{m} + \hat{n}) $$

\noindent
As indicated above, for wavenumbers larger than $\kappa_{\mbox{crit}}$, we can average over the turbulent structure, because these scales are well-sampled even for moderate exposures.  The analysis follows closely that which is typically invoked for the long exposure limit.

Specifically, we assume the Kolmogorov theory of turbulence, which posits that energy is injected into an atmospheric layer on large scales and cascades down to smaller and smaller scales via turbulent eddies, until it eventually can be dissipated by molecular viscosity.  The instability is characterized by the Reynolds number.  For typical atmospheric parameters, the Reynolds number remains above the critical value down to scales of order millimeters, while energy is injected on scales of 10 meters.  Therefore, the structure is turbulent over a very broad range of wavenumbers.

In Kolmogorov's statistical treatment (\citealt{kolmogorov}), the rate of energy per unit mass that is injected is equal to that which is dissipated down through the cascade to smaller and smaller scales.  Simple dimensional arguments then suggest that the characteristic velocities obey the relation

$$ v \sim \epsilon^{\frac{1}{3}} L^{\frac{1}{3}}$$

\noindent
which means that the temperature fluctuations are proportional to $L^{\frac{2}{3}}$.  Index of refraction variations are basically proportional to the temperature fluctuations, so they are proportional to $L^{\frac{2}{3}}$ as well.  Integrating over a layer, which is thick compared to the characteristic eddy size, leads to a two-dimensional phase power density proportional to $r^{\frac{5}{3}}$, where $r$ is a spatial scale on the two-dimensional layer.  In spatial frequency space, the power spectral density is then proportional to $\kappa^{-\frac{11}{3}}$.  The full expression for this power spectral density is

$$ \Phi_{\phi} (\kappa ) = (0.033) 2 \pi k^2 \kappa^{-\frac{11}{3}} \int_0^h dz C^2_n (z)$$

\noindent
Here $C^2_n$ is the so-called structure parameter.  It is a measure of the distribution of seeing contributions as a function of altitude.  It has units of $m^{-\frac{2}{3}}$.

We now consider $B_> ( \vec{\rho} )$.  Taking the average over the turbulence yields

$$ B_> ( \vec{\rho} ) = \frac{1}{A} \int_A d^2 \vec{r} < e^{i(\phi_> (\vec{r}) - \phi_> (\vec{r} - \vec{\rho} ))} >$$

\noindent
Expanding the complex exponential as a Taylor series, and recognizing that $\phi_> ( \vec{r} )$ is a Gaussian random variable with mean zero, it is clear that the odd moments vanish, so that $B_>$ can be written in the form

$$ B_> ( \vec{\rho} ) = e^{-\frac{1}{2} D_{\phi} ( \vec{\rho})} $$

\noindent
where $D_{\phi} ( \vec{\rho} )$ is the phase structure function, given by

$$ D_\phi ( \vec{\rho} ) = < [ \phi_> ( \vec{r}+\vec{\rho} ) + \phi_> ( \vec{r} ) ]^2 > $$

\noindent
Note that this can be written as

$$ D_\phi ( \vec{\rho} ) = 2 [ B_\phi (0) - B_\phi ( \vec{\rho} ) ]$$

\noindent
where

$$ B_\phi ( \vec{\rho} )= < \phi (\vec{r}) \phi (\vec{r}+\vec{\rho}) > $$

\noindent
is the correlation of the phase itself.  Since $B_\phi ( \vec{\rho} ) $ is the Fourier transform of the phase power spectral density, so the phase structure function can be written as an integral over spatial frequency for the turbulent layer.  Since we only want the contribution due to wavenumbers greater than $\kappa_{\mbox{crit}}$, we can truncate this integral accordingly.  The resulting expression is

$$ D_\phi ( \rho ) = 4 \pi \int_{\kappa_{\mbox{crit}}}^{\infty} [ 1 - J_0 (k \rho)] \Phi_\phi (\kappa) \kappa d \kappa $$

\noindent
The equations above yield a closed form expression for $B_> (\vec{\rho})$.   The Fourier transform of that expression gives us $PSF_> ( \hat{m} )$.

For wavenumbers smaller than $\kappa_{\mbox{crit}}$, the statistical average cannot be invoked if we hope to make a useful model of the anisotropy and decenter.  In this regime, we have to calculate the phase screens directly.  The exact approach would be to Fourier transform those phase screens, but this is computationally intensive for screens large enough for large field of views, especially when one considers the motion of the screens associated with the wind velocities during the nominal exposure.  Instead, we adopt a computationally much faster approach, which utilizes the refractive approximation.  That approximation can be justified at these large scales by the following argument.

Consider a two-dimensional turbulent patch on the phase screen of characteristic size $r$.  Radiation can interact with that patch in two related but distinct ways:  First, if there is a gradient in the index of refraction over the patch, it will act like a prism, so that the light will be refracted at a finite angle.  This displaces the image of a star, i.e. gives rise to image motion without increasing the image width.  Second, the finite size of the turbulent patch gives rise to diffraction.  This increases the width of the image, without, in general, causing significant displacement.

The two effects depend very differently on the size of the patch.  As discussed earlier, the phase power for Kolmogorov turbulence is proportional to $r^{\frac{5}{3}}$.  We show below that the image displacement due to refraction is given by

$$\delta \vec{\theta} = \frac{\lambda}{2 \pi} \vec{\nabla} \phi$$

\noindent
so $\delta \theta$ is proportional to $r^{-\frac{1}{6}}$.  For diffraction on the other hand, $\delta \theta$ is proportional to $r^{-1}$, so the former is dominant on large scales, while the latter is dominant on small scales.  The two effects are nearly exactly equal at the scale of the Fried parameter, $r_0$, which is a measure of the seeing, and is defined by the relationship

$$ r_0 = {\left( 0.423 k^2 \int_0^h dz C^2_n (z) \right)}^{-\frac{3}{5}} $$

\noindent
For good seeing sites, $\sim$ 0.6 arcsec, $r_0$ $\sim$ 10 cm or so.  Therefore on scales significantly larger than this, approximating the effect of the turbulence purely refractively is appropriate.

The argument for the refractive approximation is even stronger if the primary effects one is interested in involve PSF asymmetry and decenter.  Since a given scale is sampled multiple times over the pupil plane during a finite length observation, the net effect of image displacement due to that scale is reduced by a factor of $N^{-\frac{1}{2}}$, where N is the number of samplings.  Since $N$ is inversely proportional to $r^2$ then $N^{-\frac{1}{2}}$ is proportional to $r$, so the contribution to PSF asymmetry and decenter scales like $r^{\frac{5}{6}}$, i.e. the largest scales completely dominate.

The arguments above suggest that the natural choice for $\kappa_{crit}$ is of order $2 \pi / r_0$.  At smaller wavenumbers, refraction is dominant, and we can ignore diffraction completely.  At larger wavenumbers, diffraction is dominant, but the full effect of the turbulence can be modeled statistically.  The fact that $r_0$ is much smaller than the aperture of the telescope for modern telescopes like LSST means that this approach works extremely well for our application.  The scale of the Fried parameter is sampled many times in the pupil plane, so even invoking the refractive approximation will merely cause circularly symmetric image blur with a characteristic width of $\lambda / r_0$.  That is what a true diffraction calculation would give anyway for the contribution from this scale.

The analysis above suggests an especially simple implementation approach for an image simulation code.  The refractive approximation, formally, corresponds to an expansion of $\phi ( \vec{r} )$ as a Taylor series in $r$, and only keeping the first term

$$\phi ( \vec{r} ) \approx \phi_0 + \vec{r} \cdot \vec{\nabla} \phi + \ldots $$

\noindent
When that substitution is made, we get

$$ PSF ( \hat{n} ) = T ( \hat{n} - \frac{\lambda}{2 \pi} \vec{\nabla} \phi ) $$

\noindent
where $T ( \hat{n} ) $ is the diffraction profile of the telescope itself.  So the effect is purely image displacement.  That means that the refractive approximation can be trivially implemented using ray optics:  We bring photons individually down through atmospheric layers, and simply deflect them according to the local gradient of the phase.  This will work as long as the resolution of the phase screens does not exceed $\kappa_{\mbox{crit}}$.

This functionally accounts for the contribution  $PSF_< (\vec{m})$  As shown above, the total PSF is the convolution of the PSF’s from the lower and higher wavenumber regimes.  The convolution of two probability distributions gives the total probability of displacement if the contribution of each process is completely independent.  For $PSF_> (\vec{m})$, we have a closed form expression for this probability distribution.  We can thus sample that distribution, in both magnitude and direction, and give the photon a second deflection accordingly.  This two-kick raytrace then fully accounts for the net affect of that phase screen on all spatial scales.

Numerically, we implement the two-kick raytrace by first generating two sets of turbulent screens (4 for each layer):  the first ($\phi_< (\vec{r})$) with turbulence generated from the van Karman spectrum only with wavenumber below $\kappa_{crit}$ and the second ($\phi_> (\vec{r})$) with wavenumbers above $\kappa_{crit}$.  In practice, we found choosing $\kappa_{crit} = {2 \pi} / {1.5 r_0}$ to produce the most stable and accurate results, which is consistent with the estimate discussed above.    For the interactions of photons with the first set of screens we use the refractive approximation implemented by taking the derivative of the phase.  This interaction produces the ellipticity and image motion of the PSF.  The interaction with the second set of screens is accomplished by asymptotically averaging the aperture-weighted Fourier transform of these screens and producing a single time-averaged PSF.  This is equivalent to the long-exposure PSF for just this high frequency set of screens, and is equivalent to the closed-form solution we discussed above but in practice we use our exact generated screens.
The effect of this second set of interaction adds additional wings to the PSF as expected.  The two are implemented in sequence by performing the convolution in a Monte Carlo.  We scale the angular displacement in proportion to $\lambda^{-\frac{1}{5}}$, which is consistent with the overall scaling from complete diffraction calculations (\citealt{fried1965}).  If we choose $\kappa_{crit}$ to be an arbitrarily large number and therefore use the refractive approximation with the entire spectrum, then we get an unphysical circularly symmetric PSF which is consistent with behaviour in \cite{devries2007}, so the hybrid approach is necessary for proper treatment of the high frequency power.

  Thus in this approach we are making two approximations that are testable:  1) that the refractive approximation is valid for the first set of screens and any speckle inference structure averages sufficiently and 2) that the time-averaging of the small spatial scales is applicable.  We test this combination of these two approximations quantitatively in \S4 demonstrate its validity for reasonable exposure times $t > 1s$.  The net effect of this hybrid approach is that the computational efficiency is about four orders of magnitude larger than by using a pure Fourier approach.\footnote{A full diffraction calculation with no
approximations would involve a series of mathematical operations over every
screen pixel having size, $p \ll r_0$, for a screen area of $D v t_{exp}$ where $D$ is the
telescope aperture and $v$ is a typical wind velocity and $t_{exp}$ is the
exposure time.  Alternatively, our approximation only involves a mathematical
operation for each photon for the refractive calculation, and a one time diffraction calculation for all
sources.  Thus, the computational speed up scales as $ (D v t_{exp}
  N_{source}) \// (N_{photon} p^2)$.  For $D=10 m$, $v=10m/s$, $t_{exp}=10 s$,
$N_{photon} \// N_{source}=1000$, and $p=1 cm$, this ratio is $10^4$ and
is close to the actual observed computational ratio.}

\subsubsection{Photon Atmospheric Opacity and Dispersion Interactions}

We simulate molecular opacity by considering the wavelength-dependent optical
depth through each segment of the atmosphere layers. We use the HITRAN
database to calculate the absorption line cross-section for the water and
molecular oxygen absorption lines  (\citealt{rothman2009}, \citealt{rothman2005}, \citealt{rothman2003},
\citealt{rothman1998}, \citealt{rothman1992}, \citealt{rothman1987}, \citealt{hitran}).  We
follow the methodology described in the appendix of \cite{rothman2009} to
calculate the absorption line intensity.  We include pressure broadening
so we get a different opacity at each altitude.  We use the formulae of
\citealt{green1988} for the calculation of the cross-section due to ozone.
 The opacity of the atmosphere is determined by calculating the local
optical depth for each segment of the photons path.  Between each atmospheric
screen the column density of each molecular species of the atmosphere is
calculated by integrating the density profiles in our bulk atmosphere model and the
path length the photon.  Thus, the probability that the photon is destroyed along its particular path segment is
equal to $e^{-\tau(\lambda)}$ where

$$\tau = \sum_i \tau_i(\lambda) = \sum_i \int \sigma_i (\lambda, h)  n_i(h) dh$$

\noindent
where the sum is taken over all the species of molecules.  We also include the opacity due to aerosol scattering by including an additional contribution to the optical depth given by:

$$\tau = \tau_0 \left( \frac{\lambda}{500~\mbox{nm}} \right) ^{\Gamma}$$

\noindent
where $\tau_0=0.02$ and $\Gamma=-1.28$ are the default parameters, but vary at different sites due to a different mixture of aerosol types.

When a photon hits a cloud screen pixel it has some probability of being
absorbed (destroyed or lost by scattering outside the aperture).
We use an average opacity of 0.85 magnitudes for each of
the two clouds screens with a 1 $\sigma$ normal variation equal to the chosen
mean opacity divided by 11.3.  The latter value was chosen so that the
structure function normalization is consistent with that determined by \cite{ivezic2007}.  We
chose to make the variance proportional to the mean since the
measurements of \cite{ivezic2007} demonstrated that when the total cloud
opacity is larger, there also seems to be a proportionally larger photometric
variation.    This results in an average total opacity of 1.2 magnitudes, and has
a broad distribution, and is reasonably well matched to typical photometric
variation of actual sites.  These parameters may be different for different sites.

Atmospheric dispersion is simulated by using the index of refraction in air from \cite{filippenko1982}

$$
 1 + 10^{-6} \left( 64.328+\frac{29498.1}{146-\frac{1}{\lambda^2}}+\frac{255.4}{41-\frac{1}{\lambda^2}} \right)
     P \times $$
$$\frac{1+\left(1.049-0.0157 T\right) 10^{-6} P }{720.883
       \left(1+0.003661 T
       \right)}-10^{-6} \frac{0.0624-\frac{0.000680}{\lambda^2}}{(1+0.003661 T)W} $$

\noindent
where $P$ is the air pressure (in mmHg), $W$ is the water pressure (in mmHg),
and $T$ is the ground temperature (in C).    This shifts the angle of the photon
depending on its wavelength by a small angle. The direction of the shift is
calculated by determining the vector to the zenith.  The net atmospheric dispersion of the center of the field can be offset for a
nominal wavelength.

\subsection{Telescope/Camera Optics Simulation}

The simulation of the telescope and camera optics is performed by defining
a series of surfaces that separate the obvious volumes
composed of various media (e.g. air, silicon, glass, vacuum).  We first
describe the geometry of the telescope and camera optics, and then discuss the
photon interactions.

\subsubsection{Geometry of Surfaces and Media}

The optical elements of telescopes can often be described by cylindrically-symmetric aspheric surfaces,

$$z(r) = \frac{r^2}{R \left( 1 + \sqrt{1-(1+\kappa)\frac{r^2}{R^2}} \right)}
+\sum_{i=3}^{10} \alpha_i r^i$$

\noindent
where the height as a function of radius, $z(r)$, is expressed in terms of a
radius of curvature, $R$, a conic constant, $\kappa$, and higher order
coefficients, $\alpha_i$.  Lenses are described by two aspheric surfaces (front
and back), whereas mirrors are described by a single surface.  The
specification of these surfaces then define the regions where the glass is located.  In those regions we compute the index of refraction using the Sellmeier equation (\citealt{sellmeier}),

$$ n = \sqrt{ 1 + \frac{B_1 \lambda^2}{\lambda^2 - C_1} + \frac{B_2 \lambda^2
  }{\lambda^2-C_2} + \frac{B_3 \lambda^2}{\lambda^2 - C_3}}$$

  We specify
the detector plane by a series of rectangular devices each having a nominal
center (x,y), a number ($n_x$, $n_y$) of square pixels  having fixed pixel
size, $p$.    For the index of refraction inside the Silicon, we use

$$n=3.36 + 0.211 E + 2.79 e^{\frac{E-3.30}{0.398}}$$

\noindent
where $E$ is the energy of the photon in eV (\citealt{phillip1960}).
Additionally, the regions of the telescope/camera system that are not held at
vacuum have the refraction index of air given by \cite{filippenko1982} as described in \S2.3.
We also model the
three-dimensional spider support structure as a series of rectangular volumes.
Therefore, the combination of the series of aspheric surfaces, detector
elements, and the indices of refraction of the media uniquely specify the
configuration of the telescope/camera optics design.

An important aspect of the simulation is to perturb the optical elements and
detector segments from their nominal positions due to expected small misalignments.  With real telescopes a large
part of the image quality budget is dominated by errors from misalignments and
surface deformations of the optical elements.  This is due to either
fabrication or assembly errors, or environmental effects.  Environmental
effects include changes in temperature that misalign or deform optical
elements, gravitational vector changes due to the different pointing
orientations inducing misalignments and deformations, and other mechanical
forces including wind and seismic variations.  Modern telescopes use control
systems that attempt to correct for most of these environmental changes, but
also may induce additional misalignments and surface deformations due
to their residual inaccuracy.  To
account for all of these possible perturbations, the
simulator includes a misalignment and surface deformation for every optical
element.  The misalignments include the 6 degrees of freedom:  decenter (2),
defocus, and Euler rotations (3).  We model the surface perturbations using a
Zernike expansion up to 5th order.   Parameters describing the tolerance for each degree of freedom are input to PhoSim for any optical design. We are
currently developing an interface to input actual engineering finite element calculations for the distortions and misalignments of surfaces for the LSST system specifically.  The simulator can then choose a distorted surface based on various state variables (temperature, elevation, and actuator positions).

A tracking model perturbs the entire telescope and camera system.  The model simply perturbs the position of the photons in the reference
frame of the camera and telescope, and represents the residual tracking
errors that are expected for a nominal tracking system with parameters describing the tracking tolerance and update timescale (default of 0.1 seconds).  We have implemented a Gaussian random walk model that varies
throughout the exposure sequence.  Every 0.1 seconds, a random walk step is taken
in both elevation, azimuth, and rotation of the camera.  The mean step is
calculated so that the final RMS size of the jitter after a given exposure
time meets the expected tolerance.  Thus, the temporal spectrum is purely white up to
0.1 seconds.  Between every 0.1 seconds, the jitter is interpolated.  We also include errors in the effective exposure time of the shutter.

\subsubsection{Photon Interactions with the Optics}

As the photon is propagated through
the telescope/camera optical system, it experiences a set of photon interactions.  The first involves dome seeing.  We model
dome seeing by perturbing the photon's trajectory by an isotropic Gaussian
angular distribution equal to the expected contribution.  If the effective eddy size in the dome is sufficiently small this is a valid approximation, but if it is not then some of the dome turbulence can still be accommodated in the ground layer.  Some of the dome turbulence may have a limited drift speed depending on the airflow in the dome.

An essential calculation is to find the location of each photon hit on a given
surface.  Consider a photon at position $(x,y,z)$ with a unit vector
trajectory of $\hat{n}$.  We loop through all possible surfaces by calculating
the ray intersection distance.   In order to find the intersection of a ray
with a given surface with height, $z=f(x,y)$,
we move the ray a scalar distance, $l$, and minimize, $\delta$

$$\delta = \vline z+\hat{n}_z l - z(x+\hat{n}_x l, y+\hat{n}_y l) \vline$$

\noindent
where the surface includes the Zernike deformations.  Before
calculating the intercept we make a three dimensional Euler transformation and spatial
transformation of the photon's position and trajectory to place the photon
into the frame of the optic.
The above equation can be solved exactly ($\delta=0$) for quadratic surfaces, so we
first approximate $z(x,y)$ with a parabolic approximation, $\tilde{z}(x,y)$, and solve the equation for
$l=l_0$.  We then compute $\delta$ between the actual surface and the
position using the propagation distance, $l_0$, and choose a new value of $l$
equal to

$$l = l_0 + \frac{\delta}{\hat{n}_z}.$$

\noindent
We iteratively change $l$ using this equation until it converges to within a
tolerance of 0.01 microns.  For even highly aspheric surfaces the method
usually converges in 3 to 5 iterations, which is essential for the computational
efficiency of the code, since there are many surfaces in a typical telescope.

When the next interaction surface is chosen, the photon is moved to the new
position.  If the surface is a mirror we reflect the ray across the normal to
the mirror using a three-vector computation.  The normal has been
pre-calculated for all surfaces including the surface perturbations.  The ray's
propagation vector, $\hat{n}$ is then modified. If the surface is a lens, we
refract the ray by applying Snell's law, also using a three-vector
computation with the indices of refraction on the two sides of the surfaces determined by the
formulae discussed above.  We currently assume that the index of refraction is constant within the
material for a given wavelength, but implementation of spatial varying indices of refraction is straight-forward.
The detector elements and lens elements are treated
identically as equivalent optical surfaces.

The optical elements have an interference coating that may affect the transmission
of the photon.  The probability of transmission is a function of both
incident angle and the wavelength.  We include the
full wavelength band in the simulation, so out of band filter leaks can be properly modeled.  The
two dimensional transmission probability is calculated using a full
electro-magnetic interference boundary calculation through the actual surfaces.
When the photon reaches a coating, we use its wavelength and angle to decide
whether it is reflected or transmitted.  If the interacting surface is the
mirror, the photon is destroyed if it is not transmitted.  In the case of
filter coatings or lens and detector anti-reflective coatings, the photon can
be reflected backwards.  Thus, we use the same reflection algorithm and allow
the photon to propagate backwards.  This implies that ghost patterns are included in
the simulation.  We currently have a four column interface which accepts coating
reflection and transmission functions that are a function of both angle
and wavelength.  Coating descriptions are typically defined in this form, but
we also have an external code for the complete EM multi-layer
calculation which can calculate this format when the multi-layer structure is known.

Note that we have already included the effect of the diffraction of the
telescope pupil in the small-scale phase screen of \S2.3.2.  The
point-spread-function therefore properly includes an Airy-like component due
to the entrance pupil.  There is some freedom, however, in whether or not to imprint
the spider pattern in the Fourier calculation.  If we do that, there is a
significant variation in the projected spider size when light at large
off-axis angles is included and, in principle, every source has a slightly
different diffraction pattern.  Alternatively, we can use the edge
diffraction calculation method of \cite{freniere1999}, where the photon's
position is shifted by an angular deflection of $\lambda \// \left( 4 \pi d \right)$,
where $\lambda$ is the wavelength of the photon and $d$ is the closest
distance a photon ray gets to the edge of any part of the spider structure.
Thus, $d$ can be calculated in fully three-dimensions and this calculation
then results in both the correct geometric shadowing of the spider structure as
well as the radial envelope of the diffraction spikes, but not any interference
modulation of the diffraction spike pattern.

To represent the incoherent scattering that occurs from micro-roughness on the
mirror surfaces, we use a simple empirical model for large angle scattering.
The micro-roughness of mirrors (at the nm level) primarily causes photons to
scatter to very large angles (few arcminutes).  At the current time, we have not
implemented a physical model for this, but instead invoked an empirical radial distribution
determined for stars measured with the Gemini South telescope:

$$ \frac{1}{1+\left( \frac{r}{r_0} \right)^{n}} $$.

\noindent
We set the fraction of light in this diffuse halo compared to the core at
a fixed fraction, $f$.
Therefore, at the start of the telescope simulation the photon has a
probability, $f$, of being scattered according to the above formula.
The best fit profile had a value of $(f,n,r_0)=(0.135,3.5,0.1^{\circ})$.
There may be a contribution due to Mie scattering from dust particles included
in the above formulae (\citealt{king1971}; \citealt{roddier1995}; \citealt{racine1996}).

To determine the probability of photoelectron conversion in the silicon detectors, we
calculate the mean free path of the photon as it enters the silicon.  The
absorption coefficient depends on the device temperature and the photon
energy, $E_{\gamma}$ according to the model of \cite{rajkanan},

$$ \alpha(T)= \sum_{i=1,2;j=1,2} C_i A_i \Bigg[ \frac{\left( E_{\gamma}-E_{gj}
    (T)+E_{pi} \right)^2}{e^{\frac{E_{pi}}{kT}}-1}+$$

$$ \frac{\left(
    E_{\gamma}-E_{gj}(T)-E_{pi} \right)^2}{1-e^{-\frac{E_{pi}}{kT}}}
  \Bigg]+A_d \sqrt{E_{\gamma} -E_{gd}(T)}  $$

\noindent
where $E_g(T)$ is given by

$$E_g(T)=E_g(0)-\frac{\beta T^2}{T+\gamma}$$

\noindent
where $\beta=7.021 \times 10^{-4} eV K^{-1}$, $\gamma=1108 K$,
$E_{g1}(0)=1.1557 eV$, $E_{g2}(0)=2.5 eV$, $E_{gd}(0)=3.2 eV$, $E_{p1}=1.827
\times 10^{-2} eV$, $E_{p2}= 5.773 \times 10^{-2} eV$, $C_1=5.5$, $C_2=4.0$,
$A_1=3.231 \times 10^2 \mbox{cm}^{-1} eV^{-2}$, $A_2=7.237 \times 10^3 \mbox{cm}^{-1}
\mbox{eV}^{-2}$, and $A_d=1.052 \times 10^{6} \mbox{cm}^{-1}
\mbox{eV}^{-2}$.  The mean free path is then calculated by taking the inverse
of the absorption coefficient.  We then calculate the actual conversion path
length by taking the mean free path and multiplying by an exponentially
distributed random number.

If the conversion length exceeds the full depth of the silicon,
we allow for reflection off the backside of the device.  If
reflection occurs, we continue the conversion calculation on the reflected ray.
The reflection probability actually depends on a full EM interference
calculation that can result in fringing.   We calculate the reflection
probability using a single layer of silicon with a height that is a function
of position that depends on our perturbations using an EM single-layer calculation.  The silicon interference
probability then depends on the index of refraction on the front surface
 and on the back surface as well as the photon polarization.

We also include the possibility that the photon converts in an effective {\it
  field-free} region at the back surface.  On the back surface of
devices there may be a residual field-free region due to the manufacturing
process (such as using laser annealing).  To simulate this effect, we simply remove any electrons that have converted in the field-free
dead layer.  This then will have the correct wavelength-dependence
based on the photon conversion mean free path.

\subsubsection{Electron Interactions}

After the photon has converted to an electron, we simulate the charge
diffusion profile as it is drifted to the readout.  To do this, we have
developed a model of the electric field profile in the silicon.

$$ E_z (z) = \frac{V}{t_{Si}} + \frac{q}{\epsilon_0 \epsilon_{Si}}
\int_{t_{Si}}^z dz n_d (z) $$

\noindent
where V is the overdepletion potential, $t_{Si}$ is the silicon thickness,
$\epsilon_{Si}$ is the permittivity in silicon, and $n_d$ is the doping
density function which is given by

$$n_d(z) = n_{bulk} + n_b e^{-\frac{(t_{Si}-z)}{s_b}} + n_f e^{-\frac{z}{s_f}}$$

\noindent
Note, that the impurity density, $n_{bulk}$ is not necessarily a constant due to the difference
in segregation coefficient between the dopant and the silicon.  The impurity may have a ``tree-ring'' pattern centered on the axis of the original boule.

The relevant electron transverse diffusion at each height is calculated
with Gaussian diffusion width, $\sqrt{2 D t_c}$, where $D$ is the diffusion
coefficient, $D= \left(\mu_q(E,T) k T \right) \//q $, and the collection time is $t_c =
\int_{z_c}^{z} \frac{dz}{{\vline \mu_q(E,T) E_z(z) \vline}}$.  We have
included the effect of the velocity saturation of the electron  in the
expression for $\mu_q(E,T)$

$$ \mu_q(E,T)=\frac{1.53 \times 10^9 T^{-0.87}}{1.01 T^{1.55}
  \left(1+E/\left(1.01 T^{1.55} \right)^{\beta} \right)^{\beta}}$$

\noindent
where $\beta=2.57 \times 10^{-2} T^{0.66}$, $T$ is given in K, and $E$ is given in V/cm.  After we compute the charge
diffusion at the position in the silicon, we use the Gaussian width to move
the electron laterally in the silicon and place it at the readout surface.
Finally, the electron's position is quantized by determining in which pixel it is located.

In addition to the diffusion there is a small lateral mean shift due to any small lateral field.  These lateral field result from impurity variations, edge effects, charge stops, and accumulated charges during the exposure.  Characterizing these lateral fields in silicon devices is an active area of research (e.g. \citealt{kotov2006}).  The lateral kick the photon receives during its drift is given by

$$\Delta x=\int dz \frac{E_x(x,y,z)}{E_z(x,y,z)} $$
$$\Delta y=\int dz \frac{E_y(x,y,z)}{E_z(x,y,z)} $$

\noindent
We allow for simple parameterizations of the transverse field to add this complication.  We are continuing to evaluate the ideal parameterization based on real devices.  The lateral field is also known to  modified by the accumulated charge (\citealt{antilogus2014}).
 In addition to lateral field, real devices may not have perfectly square pixels due to lithography errors, which can be simulated by simply making a non-regular map for the pixels at the readout.

The simulation proceeds by collecting electrons in pixels.  We simulate the
effect of charge saturation and bleeding by first not allowing a given pixel
to exceed the full well depth in electrons, $w$.  Once the full well depth is
exceeded we move that electron towards the end of the row in either
direction.  We do not allow the electrons to move past the implant at the
center of the device, and place the electron in the closest unfilled pixel
along that row.  Once the entire row has exceeded the full well depth, we
remove the electron entirely.  This then approximates the effect of bleeding.

Actual images of cosmic rays are added to the simulated images using real data from
thick silicon devices \cite{doty}.  To do this, we constructed postage stamp images of 130 different
actual cosmic ray events.  We then add these randomly to our simulated images
using two important calculations.  To determine how often to place a
cosmic ray in the image, we use the production rate of 0.04 cosmic rays per
sq. centimeter of silicon per second.   The actual cosmic rays are expected to
be a combination of gamma rays from local ground radiation and muons and other
particles from atmospheric particle interactions.  Our data have some
combination of the two, but perhaps not in the correct proportions.  A second
calculation gives us the scaling of electrons in the cosmic ray data, to the
appropriate volume of silicon of the simulated pixels.  This correctly normalizes the correct number of ionized
electrons in the simulated devices.

\subsection{Electronic Readout Simulation}

Hot pixels are added to the image by randomly choosing a fraction of the
pixels and then placing electrons equal to the full well depth.  Similarly,
a fraction of the pixels are flagged as dead and then the electrons are
removed from those pixels.  Hot columns are simulated by selecting some
fraction of pixels that are the ends of hot columns and then setting those
pixels and the pixels behind that pixel in the readout chain to full well
depth.  Dark current is computed for the length of exposure, and modeled by
randomly adding a number of electrons to each pixel with Gaussian error.

The CCDs are segmented according to the amplifier readout scheme.
Rows or columns are added according to the number of pre-scan or over-scan
pixels.  Each pixel is then assigned a readout sequence according to the
parallel and serial charge transfers.  Electrons then have some probability of
being shifted to a pixel behind it in either the serial or parallel direction during
the readout.  We found that it was necessary to perform a shift for
every pixel individually, since the CTE values can be quite high with
modern devices ($>99.999\%$), and it is not possible to make a multinomial
approximation for this effect.  We then loop through the
pixels in readout order.  Read noise is implemented by using a Gaussian with mean equal to the expected
read noise value.  We also vary this value between amplifiers.

Finally, the digitization process can be approximated by

$$ ADU = \frac{e}{G \left(1+N \frac{e}{W} \right)} + B$$

\noindent
where $e$ is the number of electrons in a given pixel, $G$ is the gain, $W$ is the full well
depth, $B$ is the bias, and $N$ is a non-linearity factor.  We have not
implemented a detailed model of gain and bias variations across each segment or as a function of time,
but we do vary these parameters between amplifiers.  We also have implemented possible digitization errors
in which each bit is modified by

$$ {\mbox bit_i}=\left( \frac{ADU}{2^i}  +{\sigma_i} \right)~~{\mbox mod}~~2$$

\noindent
and then the final ADU value is given by

$$ADU = \sum_i bit_i 2^i$$

\noindent
where the $\sigma_i$ values are empirically determined estimates that keep the digitization from being perfect.

\section{Software Implementation}

The software is constructed to simulate a series of images in identical form
to what a real telescope would produce.  The input resembles the combination
of operational commands a real telescope operator would execute, and a set of
astrophysical catalogs.  The final output is a sequence of FITS images
produced from individual amplifier chains of CCD devices.  Following, we
discuss the overall architecture and numerical implementation of the physics
we described in \S2.

\subsection{Architecture}

PhoSim is written in object-oriented C++ code.  The
codes are run with Python scripting.  The overall architecture is
shown in Figure 1.  The C++ code is divided into 5 parts:  the atmosphere
creator, the instrument configuration, the trim program, the photon raytrace,
and the electron to ADC codes.   The codes are configured to simulate a
particular visit (a series of exposures at a given place on the sky).   The
first two codes set up all the input data that
are required to describe the current atmosphere and instrument configuration.
The trim program is then run for every chip where the astrophysical catalog is
reduced to only objects that have a significant chance of producing photons on
that particular chip (either object centered in the projected sky tile for
that chip or particular bright objects that may have a large scattering halo
or scattered ghost photons).  This facilitates parallelizing the calculation.
The photon raytrace simulates the
individual photons through the atmosphere, telescope, and camera and collects
the converted photo-electrons in an image.  Finally, the electron readout is
simulated and the image is digitized in the electron to ADC code.

The input and output to the code are well defined.  The input consists of an
{\it instance catalog}, which is a list of objects in the sky at the
particular time of the observation and a description of all the properties
needed in \S2.2.  The instance catalog also includes the commands that a
telescope operator would have available and other environmental parameters
that may affect how the observation would be done.  Other astrophysical
information, such as the position of the Sun and Moon is included as well.  We
also have a physics command file as an optional second input.  This includes
any commands to override our representation of the most realistic physics.
This can be used in a large number of ways including by turning off a subset
of effects in a modular way or setting parameters to specific values.
Those are useful options, both for validation and testing, and also for studying the physics
that might lead to a systematic error in a particular image processing algorithm.

The main output of the code is raw digitized FITS images for every amplifier
on every CCD.  There are alternative other outputs
as well that are unavailable with a real telescope.  We can output an event
file, which describes the interaction position of every photon as it is
propagated through each layer or surface.  We also can output the actual
number of photons detected from every source and the mean coordinate of those
photons.  This information would be only approximately known from the images.
We also can output the relative throughput of photons at each layer or optical
surface.

The PhoSim code only has two package dependencies:  cfitsio
(\citealt{pence1999}) and fftw3 (\citealt{frigo2005}), and is otherwise built
using both standard C/C++ libraries and custom numerical codes.  This allows us
greater custom numerical detail, and makes the installation straightforward.
The entire phosim code is designed so it can be implemented easily on grid
computing.  The architecture was designed so that I/O would be minimal and the
simulations could be done in parallel at the chip level.  The package can be
run with both a script for laptop/desktop simulations and another script that
uses CONDOR to generically run simulations on grid computing systems (\citealt{condor}).

We have built an extensive validation framework with the PhoSim code.  The
framework includes both unit testing and integration testing.  The unit
testing executes individual functions to assess whether the return values
obtain the correct values.  Integration tests run the entire suite of code and
determine whether measured properties of images obtain measured values within
a specified tolerance.  Integration tests use instance catalogs of a limited
number of objects (usually stars and galaxies), and then use the configuration
files to run the photon simulator in a variety of configurations.  We describe
some of the results of the validation tests in \S4.

The entire Photon Simulation package is on a Git repository that is
available at https://www.bitbucket.org/phosim/phosim\_release.  Associated documentation is available at this site.
We release a tagged version several times per year.  We use a modern modular
object-oriented software design approach where the speed of the code is a very
important consideration that we discuss in the following section.

\subsection{Numerics and Optimizations}

There are a variety of numerical implementation details specific to each of the implemented physical effects.
In general, Monte Carlo simulation times are
proportional to the number of points in the Monte Carlo integration (in our
case, the numbers of photons) and a fixed overhead associated with setup.  Either reducing the time per photon or
reducing the number photons that need to be simulated can minimize the
simulation time.  For the former, minimizing the number of mathematical operations done on each
photon reduces simulation time.  Removing redundant calculations by saving values in pre-calculated tables
wherever possible is one key to optimization.  We do this for the
shapes of the optical surfaces, transmission curves, turbulence screens,
optical depths, etc.  The overall reduction of the number of lines of code
that need to be executed in the inner loops is also important, so this has
been a priority throughout its development.  Currently, we are simulating a
photon in about 2 $\mu s$ on a 2.5 GHz processor, implying a very efficient number of calculations per photon.
This can possibly be improved further (and has during the recent development),
but it is a fairly small number of operations considering the detailed physics
and fidelity constraints.  We profile the code periodically, and the various
parts of the calculation described in \S2 contribute roughly equally at the
current time since obvious bottlenecks have been removed.

We have performed another set of optimizations that reduce the overall numbers of
photons that need to be simulated.  These optimizations have a minimal effect
on the fidelity, and all the optimization can be turned off for detailed
comparisons.  We have three such optimizations that we
call:  1) dynamic transmission optimization, 2) saturated object simulation,
3) and background optimization simulation.  Dynamic transmission optimization
works by attempting to guess whether the given photon has any chance of
surviving the Monte Carlo simulation.  The most common way for a photon to be
removed is by not surviving the filter
transmission.  A simple optimization is to pre-roll the random numbers at the
start of the simulation of the photon.  We then store a worst-case transmission
curve for each atmospheric layer and surface coatings for each wavelength as
the simulation is running.  This is necessary because of the complication that
our transmission functions are not only wavelength-dependent but they may be
angle-dependent and time-dependent as well.  Therefore, we simply estimate first
if the photon has any chance of surviving the series of transmission functions before
the photon is simulated through the full physics.  This improves the
simulation speed by an order of magnitude, and has a negligible change on the
photometry.

A second optimization is the saturated object simulation optimization.  The
brightest stars in a typical field contain a very large fraction of all
the photons.  However, the vast majority of those photons end up in a saturated
bleed trail that has a negligible amount of useful information.  To take
advantage of this, we have built an optimization that makes the ray represent
N photons instead of just one when it is highly likely that ray will end up in a
saturated bleed trail.  This can be done at run-time as soon as the pixels start to
saturate and exceed the full well depth for a given source.  To simulate
the wings of the saturated source accurately with the right statistical
properties, we are able to keep track of the rays that might end up in the
wing from large angle scattering (either diffraction spikes, mirror surface microroughness, or
diffraction in the atmosphere).   We can then enhance the probability of those
events happening by artificially looping over that physics $N$ times.  We then
choose $N$ such that the probability of the photon being kicked by an
angular distance greater than the current minimum radius of the saturation
pattern, $r_0$, is greater than a value, $\alpha$.  Normally, $\alpha$ would
be a small quantity (few percent) without optimization, but here we enhance
its probability to $90\%$.  We preserve probabilities, however, by letting the
ray represent $M$ photons, if it is not a large angle photon.  $M$ is given by

$$ M = \frac{N-\alpha}{1-\alpha}$$

\noindent
On the other hand, we only let the ray represent one photon if it is a large
angle photon. Similarly, we conserve the photon detection probability by
letting the ray represent $N$ photons, if it is removed.  Thus, the algorithm
still simulates one photon at a time for virtually all photons that will be
measurable in the image (the wings), but simulates several at a time for photons in
the saturated bleed trail or those not detected at all.

The final optimization involves the simulation of the background photons from
airglow, scattered moonlight, twilight, or dome light simulations.  Since
these photons outnumber the astrophysical photons, it is important to simulate
them efficiently.  On the other hand, we found that using parametric models
of the background (i.e. trying to predict the flux in each pixel and then
adding noise) did not result in high enough accuracy for some of the more
subtle physics details.  An algorithm that facilitates faster simulations is
to represent the ray as $N$ photons for part of the simulation.
When the photon is closer to the pupil plane, a diffuse illumination pattern
will produce an nearly identical photometric response for small angular
distances.  Equivalently, most of the differences in background from one pixel to the
next occur because of physical effects near the image plane.  Therefore, when
the photon reaches the detector we can randomly spread the $N$ photons in a
Gaussian pattern (with $\sigma$ of several arcseconds width) and simulate the detector
physics one photon at a time.  To not induce fluctuation patterns, $N$ should
be chosen so that we do not simulate many more than $\sqrt{p}$ where $p$ is the
photons per pixel.    This results in two orders of magnitude faster
background simulation, and varying degrees of accuracy depending on the choice
of $N$ and $\sigma$.  With the default parameters, a reduced $\chi^2$ of 1.1 is
obtained for comparing the images of off-axis chip simulation with this optimization turned on and off.  For detailed studies of either bright stars or background models the optimizations can be turned off so photons are properly simulated one photon at a time for complete accuracy.

\section{Results}

We implemented the LSST design details through a series of data input files.
The physics code is
written deliberately without including any reference to LSST-specific data,
so implementation of other telescopes is
straight-forward through an alternative set of design files.  We do not
describe the detailed design parameters of LSST here, but Table 1 lists the
variety of design information that is needed to describe a telescope and site.  {The physics implemented above should be appropriate for most optical survey telescopes without significant modification.  There may be possible extensions to other specialized systems as well, such as an adaptive optics telescope, where the effect of the AO control loop on the residual phase error would have to be considered.  An application to a space-based telescope would be straight-forward as well.  Extending the physics wavelength coverage into the UV and further into the IR would also require little modification.  However, the validation studies we pursue below are most appropriate for simulating optical survey telescopes.}

\begin{deluxetable}{ll}
\tablecolumns{2}
\tablecaption{Required Instrument and Site Specific Data}
\tablehead{ \colhead{Type of Data} & \colhead{Information}}
\startdata
Optical prescription for & type of optic, $R_{curv}$ \\
every optical element &  position, outer/inner radii, $\kappa$, \\
 & $\alpha_3$, $\alpha_4$, $\alpha_5$ ,
$\alpha_6$ , $\alpha_7$, $\alpha_8$, $\alpha_9$, $\alpha_{10}$
\\
& medium adjacent to surface \\ \hline
Focal plane geometry  & position, pixel size, pixel \#'s, \\
for every device &  thickness, readout speed, \\
&  shape distortion parameters \\ \hline
Surface perturbation & distribution parameters \\
and misalignments for & \\
every degree of freedom & \\ \hline
Coating for & transmission probability \\
every optical surface & vs. angle and wavelength \\ \hline
Device data & hot/dead pixel rate, \\
for every sensor &  hot column rate\\ \hline
Readout Data & amplifier segmentation, \\
for every amplifier & pre/over-scans, read noise, \\
 &   serial and parallel CTE, gain, \\
 &   bias, non-linearity \\ \hline
Site data  & turbulence distribution \\
for every layer  & outer scale distribution \\
 & wind vector distribution \\ \hline
Location & latitude, longitude, altitude\enddata
\end{deluxetable}

In Figure~\ref{fig:label2}, we show the path of the photon through the large
dynamic range of scales in the photon simulation.  The top left image shows
the path of photon in a cylindrical column through the atmosphere, the bottom
left and bottom center images show the photons moving through the telescope
system, and the bottom right image shows the conversion of photo-electrons in
the silicon.  The top image then shows the resulting images of stars and galaxies are
they are collected in pixels.  This figure does not completely show the
physics detail in the simulation.  Figure~\ref{fig:label3}, however,
demonstrates the physics detail by simulating a single star and
successively turning on more physical effects in the simulation.  Each separate part of
the simulation contributes in different ways to the size and shape of the PSF,
the photometric intensity, and the astrometric position of the image.  The
star was simulated in the u,i,y filters and combined to be a three-color RGB
image, so that the chromatic effects can be seen.  Figure~\ref{fig:label4}
shows a simulation of a star through the same optics and atmosphere configuration,
but at a point 1 degree away from the center of the field.  Through comparison of these
two figures, the subtle spatial dependence of the PSF and photometric
properties can be seen.   Table 2 summarizes the
relevant physical effects that determine the particular image properties.
Figure~\ref{fig:label5} is a collection of 3 amplifier images with various
stars and galaxies using catalogs generated by \cite{connolly}.
Every photon has been simulated and sampled from the spectral energy
distributions and spatial models in the catalog.

\begin{deluxetable}{ll}
\tablecolumns{2}
\tablecaption{Connection between relevant physics and image properties}
\tablehead{
\colhead{Image Property} & \colhead{Most Relevant Physics}}
\startdata
Photometric Acceptance & Geometric Acceptance, Coatings, \\
& Photo-electric Conversion, \\
& Atmosphere Opacity, Clouds \\  \hline
PSF size & Optical Design \\
& Atmospheric Turbulence \\
& Perturbations and Misalignments, \\
& Dome Seeing, Tracking \\
& Charge Diffusion \\
& Mirror Micro-roughness \\ \hline
Astrometric Scale & Optical Design, \\
& Atmospheric Dispersion, \\
&  Tracking, \\
 & Perturbations and Misalignments \\ \hline
Background Level & Diffuse Airglow, \\
&  Reflected Moonlight, \\
& Geometric Acceptance, \\
& Coatings, \\
& Photo-electric Conversion, \\
& Atmosphere Opacity, Clouds \\ \hline
PSF shape and Variation & Atmosphere Turbulence, \\
& Optical Design, \\
&  Lateral Detector Fields, \\
& Atmospheric Dispersion, \\
& Misalignments and Perturbations \\ \hline
Differential Astrometry & Atmosphere Turbulence, \\
&  Atmospheric Dispersion, \\
&  Lateral Detector Fields, \\ \hline
Photometric Variation & Coating Non-Uniformity, \\
& Lateral Detector Fields \\
& Interference non-uniformity \\ \hline
Background Variation & Rayleigh and Mie Scattering, \\
& Airglow Variation \\
\enddata
\end{deluxetable}

To date, we have validated the most critical aspects of the simulator.  However, a
complete validation is beyond the scope of this work and awaits wider community
involvement using real data from astronomical telescopes.  Following, we discuss the most critical
aspects of addressing any possible {\it calculational} errors (i.e. given an ideal setup can PhoSim
properly calculate a set of given quantities).  The larger future validation, however, involves
possible {\it representational} errors (i.e. does the simulator accurately reproduce the image
properties from a given telescope and site).  Representational errors are much harder to address,
since they involve understanding the exact characteristics of the telescope, camera, and site and not just
the correct implementation of the physics of photon and electron propagation.

The required calculational accuracy of PhoSim is a complex topic.  As described in the previous sections,
we avoid making approximations, unless they result in orders of magnitude
faster simulation rates.  Thus our goal is to make the physics as complete and accurate as possible.  Although there is a vast
range of science applications that would place disparate requirements on simulation accuracy, a first
order estimate of required accuracy can be obtained by considering the statistical error for measuring
various attributes of the point spread function.
 If the calculational error is significantly below the statistical error for a source with a given number
of photons then the error will be unobservable.  In the Appendix, we estimate generically that in an optical
 survey the brightest new source will have statistical photometric error of $10$ millimags, a FWHM uncertainty of
 $12$ milliarcseconds, a centroid uncertainty of $7$ milliarcseconds, and an ellipticity uncertainty of 1.4\%.  Below,
we show that the known calculational errors of PhoSim are significantly below those thresholds.

For the calculational validation of the atmosphere simulation, we show typical phase screens in
Figure~\ref{fig:label9}.  Three examples of 50m by 40m phase screens of the combined 7 layer set of screens after combining the screens on the 4 different pixel scales are shown.  Despite the complexity in constructing these images, they qualitatively
resemble phase screens generated using standard techniques (\citealt{lane}, \citealt{ellerbroek}, \citealt{lelouarn}, \citealt{britton}, \citealt{jolissaint}), but we have a representation of the phase over several kilometers.  A quantitative validation is shown in the upper left panel of Figure~\ref{fig:label9} where we calculate the two dimensional structure function
of the phase screen and compare with the analytic calculation of \cite{fried1965} for a pure Kolmogorov spectrum and the numerical integral of \cite{lucke2007} for various values of the outer scale for a von Karman spectrum.
No visible artifacts can be seen in the structure function. The propagation of light through the phase screens uses the hybrid propagation technique that can be compared with a traditional Fourier approach.  This is shown
in Figures \ref{fig:label10} and~\ref{fig:label11}.  Figure~\ref{fig:label10} shows an instantaneous exposure, whereas Figure~\ref{fig:label11} shows a typical 15 second LSST exposure.  The PSF using both numerical approaches is calculated and compared on LSST pixel scale (bottom) and finer scale (top).  The residual subtracted PSFs (right)
 show no obvious biases or errors.  Quantitative comparisons can be made by measuring weighted moments and computing the centroid and ellipticity.  Even for bright stars, the statistical errors in the centroid and ellipticity are statistically consistent using the two approaches.
  To quantify this, we performed two dozen simulations of stars using the two approaches and plot the measured ellipticities and centroids demonstrating a high correlation in Figure~\ref{fig:label12}.  The differences are shown in the histogram in the right panel and are consistent within statistical errors.  This histogram can be used to set an
upper limit to the error of the technique of 0.60 milliarcseconds for the centroid error and 0.0046 for the ellipticty error.  Thus, the only significant error of this light propagation technique may be for arbitrarily short exposures with diffraction limited telescopes where the hybrid technique will not reproduce the speckles in
Figure~\ref{fig:label10}.  However, as we discussed in \S1.3, this is not relevant for optical survey telescopes.

For the calculational validation of the instrument simulation, we show a spot diagram simulation in Figure~\ref{fig:label13}.  A spot diagram is produced in PhoSim by turning off detector effects, ignoring perturbations and misalignments of the optics, and by making arbitrarily small pixels.  We compare the result of 5 different monochromatic
 simulations for a 1.4 degree off-axis point with the commercial raytrace code, \cite{zemax}.
The rms spot size is about 0.1 arcsecond, so smaller than a single pixel.  The x-axis in Figure~\ref{fig:label13} subtracts off the large positional offset expected for such an off-axis source of 254911.0 $\mu m$.  This demonstrates that we are correctly predicting the positions of photons at least to the 7th digit.  A detailed ray-by-ray comparison
can be used to assess the quantative accuracy of the geometric raytrace.  We find that the discrepancy in ray positions has an average displacement of 0.18$\mu m$, most likely due to the numerical accuracy with which we store surface maps.  Note that this is less than 1/50th of an LSST pixel.

For the verification of photon throughput, we simulate a top hat spectral energy distribution, and
verify the analytic prediction for the number of photons. We then compare the detected
number of photons with the analytic calculation

$$ \mbox{photons}= \frac{1}{h} 10^{\frac{m_{500}+48.6}{-2.5}} t_{\mbox{exp}}
\pi \left( r_o^2- r_i^2 \right) \frac{\left( \lambda_2 - \lambda_1 \right)}{500~\mbox{nm}}$$

\noindent
where $h$ is Planck's constant, $m_{500}$ is the AB magnitude at 500 nm,
$t_{\mbox{exp}}$ is the exposure time, $r_o$ and $r_i$ are the inner and outer
radii of the aperture, and $\lambda_2$ and $\lambda_1$ are the wavelength
limits of the top hat.  Figure~\ref{fig:label14} shows the number of
photons generated using PhoSim as compared with the analytic prediction using
square spectral energy distributions, the aperture of LSST, and the exposure time.  The results are
consistent within statistical errors, and well below any reasonable science
application with PhoSim since the mean inaccuracy is 0.45 millimags ($0.045\%$).  Thus, the numerical simulation of the atmosphere and instrument is sufficient for nearly all practical science cases.  There is still considerable work to validate representational errors.

Figure~\ref{fig:label15} shows the speed of the calculation, and demonstrates the efficiency of the photon Monte Carlo approach.  The speed is
proportional to the number of photons for unsaturated sources (above 15th
magnitude) and is approximately 400,000 photons/s on a typical workstation (2.5 Ghz Mac Intel Processor).
Background photon simulations are considerably faster (factor of 60), and
saturated sources are as well (factor of 8 for a 12th magnitude star) due to
the optimization described in \S3.2.  The computation takes about 3 hours for
a single chip of LSST (13 by 13 arcminutes) filled with a typical distribution
of stars and galaxies for a 15 second exposure.  Thus, despite the physics complexity pursued in this work, we can simulate typical individual sources which may have thousands of photons in a few milliseconds.  Using large scale computing, we have run the code on about 2000 processors simultaneously for a variety of data challenges to simulate LSST simulations for the data management team.  Therefore, during these runs we are already generating data within reach of the actual real time image production rates of the future highest data rate telescope (LSST).\footnote{Since it takes a few hours to generate a single chip image of an LSST 15 second exposure, the real time ratio is $\frac{15}{3*3600} \frac{N_p}{N_c}$ where $N_c$ is the number of chips in the focal plane (189) and $N_p$ is the number of processors.}  Since the simulations are done in parallel, there is no barrier to scaling to arbitrarily large numbers of cores.

\section{Conclusion and Future Work}

We have demonstrated the implementation of atmospheric and instrumental physical effects appropriate
to astronomical image simulation for optical survey telescopes.  The
simulation approach using a photon Monte Carlo is both efficient and flexible.
Simulations with this level of physics detail are remarkably
fast with the photon Monte Carlo methodology.  Detailed second order
image properties can be studied in analyses that require high precision
(e.g. astrometry, detailed PSF modeling, photometric calibration) using this
simulation tool.

There are a variety of ways to improve and extend the code further.  Future work
involves validating physics techniques and  adding more details to the representation of the instrument and site characteristics.  The most important fidelity improvements are: 1) a physical model
predicting the perturbations and misalignments, 2) a description of lateral charge diffusion associated with
electric field distortions, and 3) more accurate atmosphere site models for both turbulence and opacity.
Detailed validation may be achieved by greater community involvement with the simulations of existing telescopes in various configurations, an
we expect that this tool will be useful for many different science applications.

\acknowledgments
JRP acknowledges support from Purdue University, the Department of Energy (DE-SC00099223),
National Science Foundations Research Experience for Undergraduates (REU) for Purdue Physics, the LSST Project (C44054L), and large-scale computing support from the Rosen Center for Advance Computing (RCAC) at Purdue University.  We thank the LSST internal reviewers, J. Thaler, M. Jarvis, \& D. Kirkby, for helpful comments and a careful reading of this manuscript.  We thank an anonymous referee for very helpful comments that improved this manuscript.

\newpage

\appendix
\section{Statistical Error of an Object's Attributes in an Optical Survey}

To determine the statistical error on a source's attributes, we simulated a ellipsoidal gaussian with $N$ photons with PSF of size, $\sigma$.  We then measured the source's attributes (FWHM, ellipticity, flux, and centroid) by a weighted second moment method.  We confirmed the scalings that the statistical error for the photometric flux is $N^{-0.5}$, the error on the FWHM is $2 \left( 2 \log{2}\right)^{0.5} N^{-0.5} \sigma$, the error on the centroid is $2^{0.5} N^{-0.5} \sigma$ and the error on the PSF ellipticity is $2^{0.5} N^{-0.5}$.

Science applications can be done with objects of all brightnesses and therefore values of photons, $N$, but
note that for an optical survey most of the interesting new sources will be near the detection threshold.
This occurs when

$$N F > s \sqrt{F p b}$$

\noindent
where $N$ is the number of photons from a source in a single frame, $F$ is the number of co-added frames,
$p$ is the effective number of pixels the source is distributed over, $b$ is the background rate per pixel per frame,
and $s$ is the source signal to noise.  Generically, $p$ is around $10$ to sample the PSF but not reduce sensitivity,
$b$ is in the range from 10 to $10^3$ to be sky-noise limited but not limit the dynamic range of a CCD up to
the full well depth (typically $10^5$ with modern devices), and a reasonable detection threshold would have
$s$ of 10.  Therefore, $N$ would typically be between $100 F^{-0.5}$ and $1000 F^{-0.5}$ for a source at the
survey's detection threshold.  Conservatively, if we consider sources an order of magnitude brighter than the
detection threshold, then the brightest new source in a survey would then have around $10^4 F^{-0.5}$ counts in a single exposure
or $10^4 F^{0.5}$ counts in the co-added exposures.  For seeing of 0.5'',
the statistical uncertainty of the object's attributes in a single exposure would have a photometric error of $10$ millimags, a FWHM of
 $12$ milliarcseconds, a centroid error of $7$ milliarcseconds, and an ellipticity error of 1.4\%.  Therefore,
a reasonable set of requirements for the accuracy of an optical survey simulator is that any known calculational inaccuracy is below several millimags of photometric error, several milliarcseconds of FWHM and centroid error, and percent-level ellipticity error.  This would be sufficient to not add a significant systematic simulation error on an object's attribute for a source with less than $10^4$ photons in a single exposure or $10^4$ combined photons in a co-added exposure.  In most cases, the simulation error would not occur in the same way in a series of simulated exposures so the single epoch requirement would be sufficient, but if it did, the error would, for example, result in a factor of 10 stricter requirements for 100 frames.  In either case, the scalings in this appendix can be used to estimate the errors on a given object that is the study of a science case.


\newpage

\begin{figure*}[htb]
\begin{center}
\includegraphics[width=0.9\columnwidth]{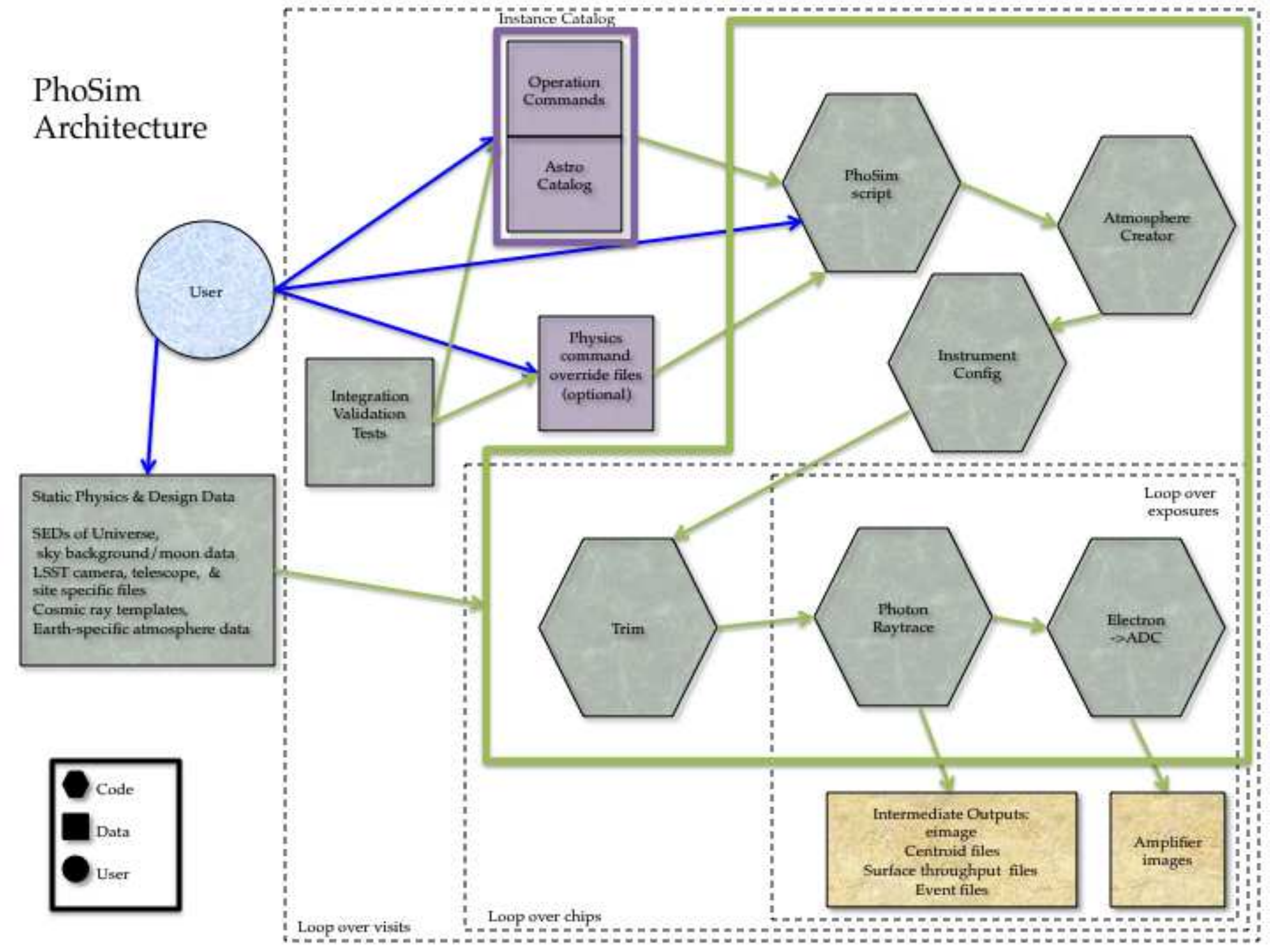}
\end{center}
\caption{\label{fig:label1}  The basic architecture of the phosim code.  The
  five codes and the controlling scripts are shown as green hexagons.  The
  inputs are the instance catalog, optional physics command files, and the
  static design data.  The user can interact with the code running the PhoSim
  script and either construct instance catalogs, modify the physics commands,
  or change the input data.  The integration validation tests are a
  combination of simple instance catalogs and physics commands.  The main
  output is amplifier images, but a number of intermediate outputs are also available.}
\end{figure*}

\begin{figure*}[htb]
\begin{center}
\includegraphics[width=0.9\columnwidth]{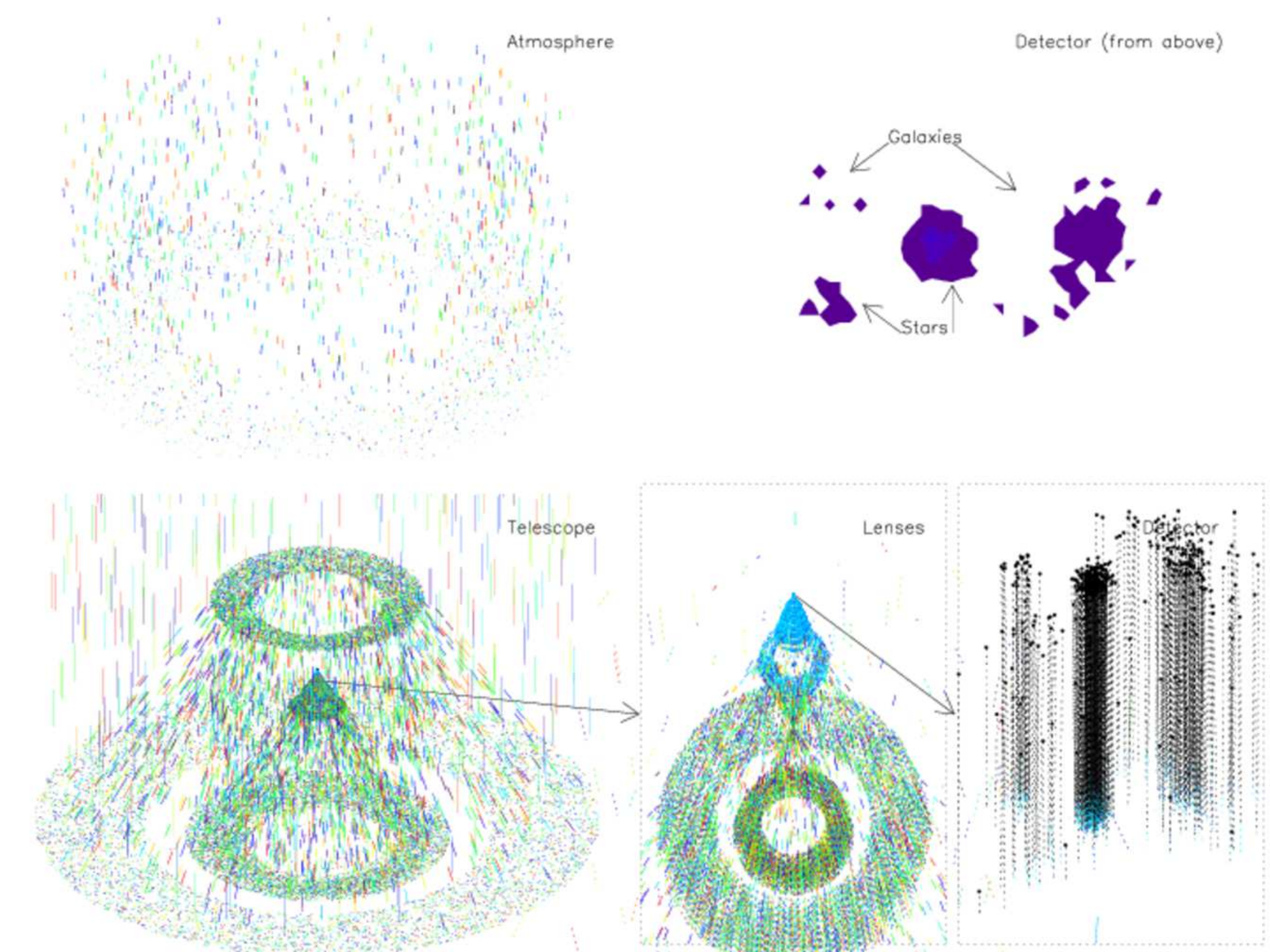}
\end{center}
\caption{\label{fig:label2}  A graphical representation of the photon's
  trajectories.  The color of the photons is proportional to the photon
  wavelength.  The upper left panel shows the path of photon's through a
  cylindrical column of the atmosphere.  The bottom left panel shows the 3
  mirror LSST design.  The bottom middle panel zooms in around the lenses and
  shows the non-blue photons being reflected backwards from the filter.  The
  bottom right panel shows the photons (blue) converting into photo-electrons
  (black) as they drift to the readout.  The top panel is a two dimensional histogram (image) of the
  final electron positions.}
\end{figure*}

\begin{figure*}[htb]
\begin{center}
\includegraphics[width=0.9\columnwidth]{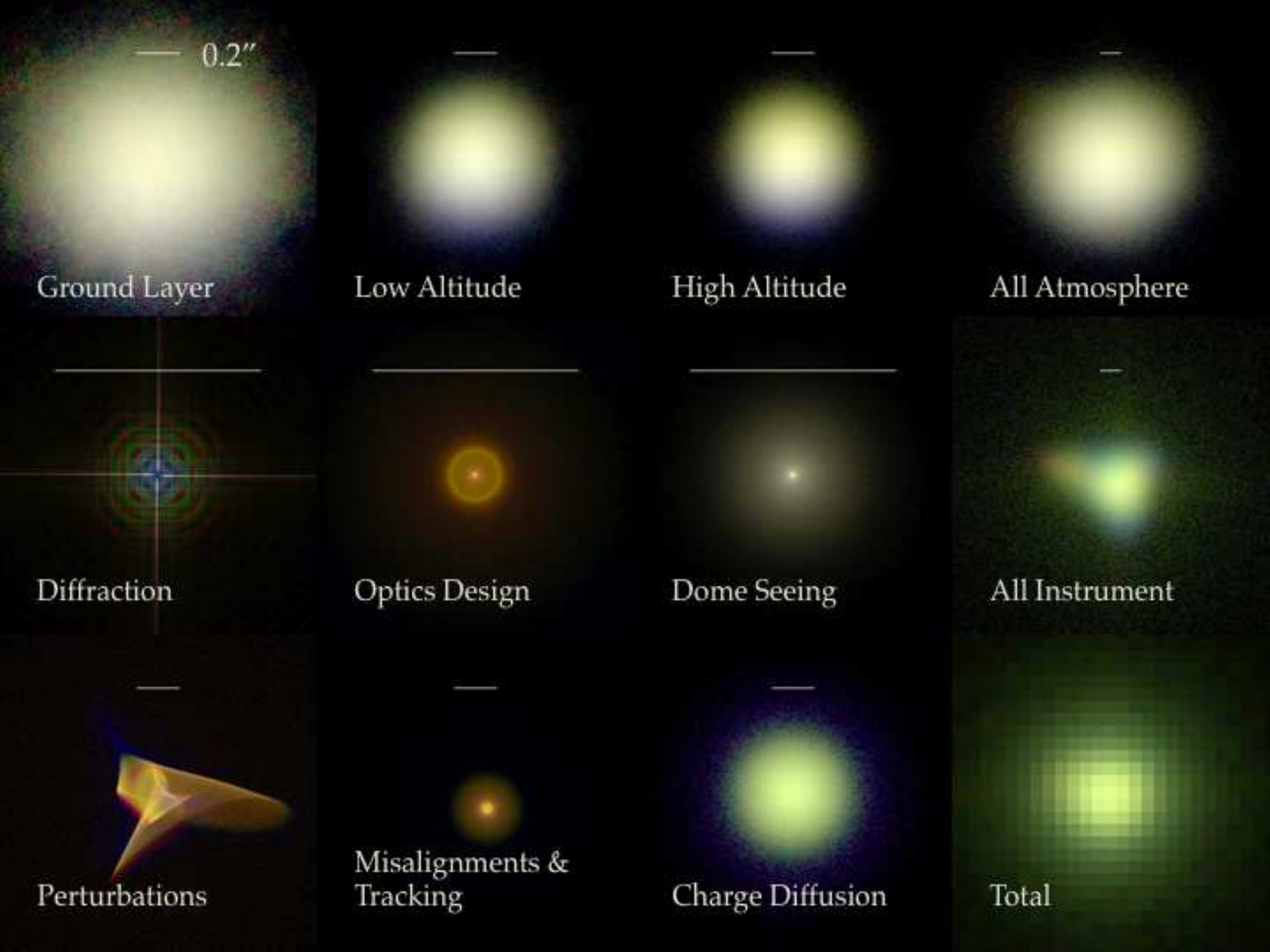}
\end{center}
\caption{\label{fig:label3}  A simulation of a single star at the center of
  the field with different physics modules turned on in each of the 12 panels.
The scale is different in each of the panels and is indicated by a bar
representing 0.2 arcseconds.  The
simulation is done in three filters (u, i, y) through the same atmosphere and
instrument configuration to show the chromatic effects of each physical
effect.  The PSF size/shape, photometric intensity, and astrometric position
is therefore changed by each part of the simulation.  The instrumental PSF is
dominated by the perturbations, misalignments, and charge diffusion.  The
atmospheric PSF is produced by a combination of the ground layer and the free
atmosphere seeing.}
\end{figure*}

\begin{figure*}[htb]
\begin{center}
\includegraphics[width=0.9\columnwidth]{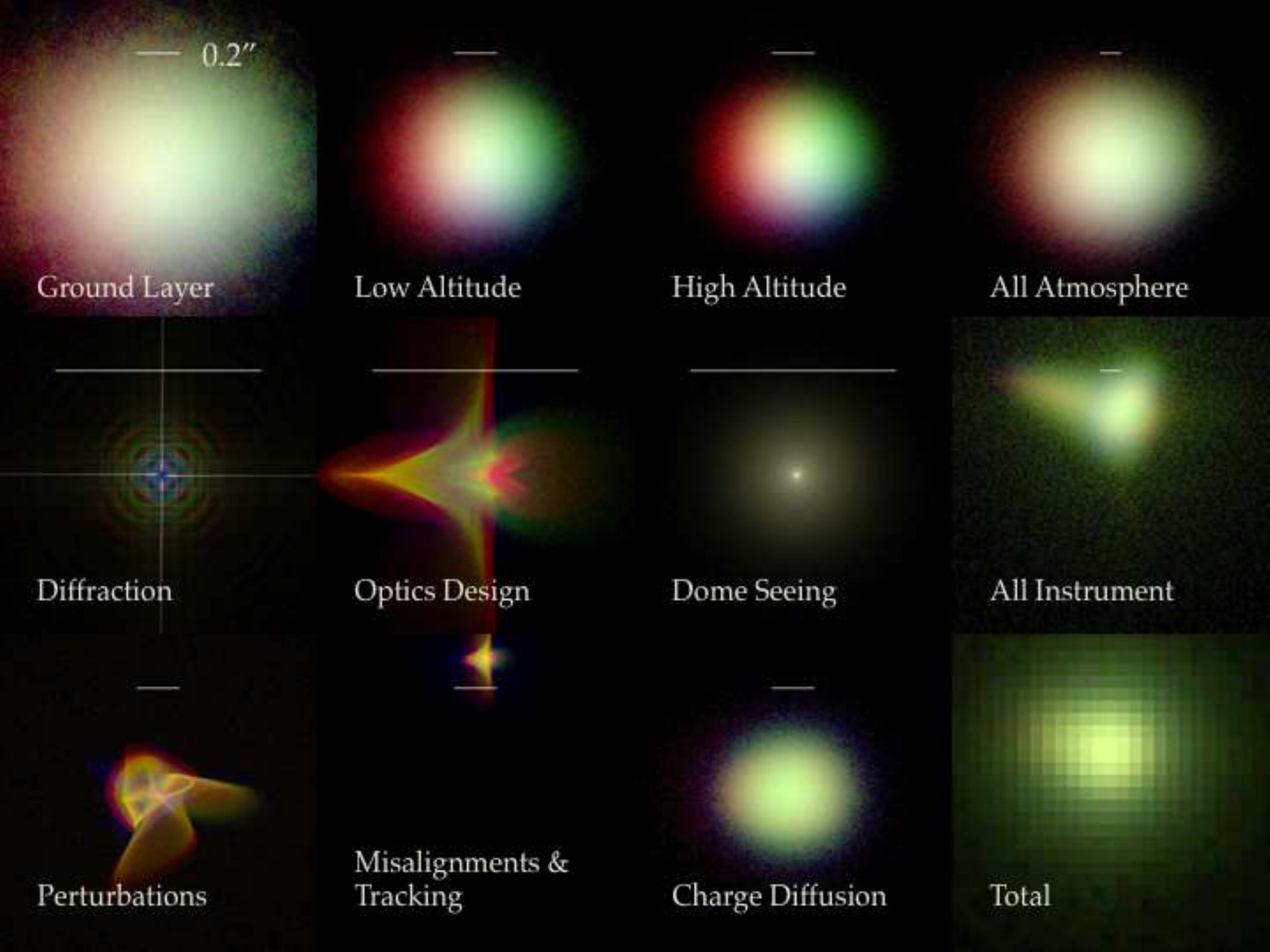}
\end{center}
\caption{\label{fig:label4}  The same as the previous figure but with the star
1.4 degrees off-axis.  Comparison of the two figures shows not only the
complex nature of the simulation, but also the prediction of the
complex field-dependence of these effects.}
\end{figure*}

\begin{figure*}[htb]
\begin{center}
\includegraphics[width=0.9\columnwidth]{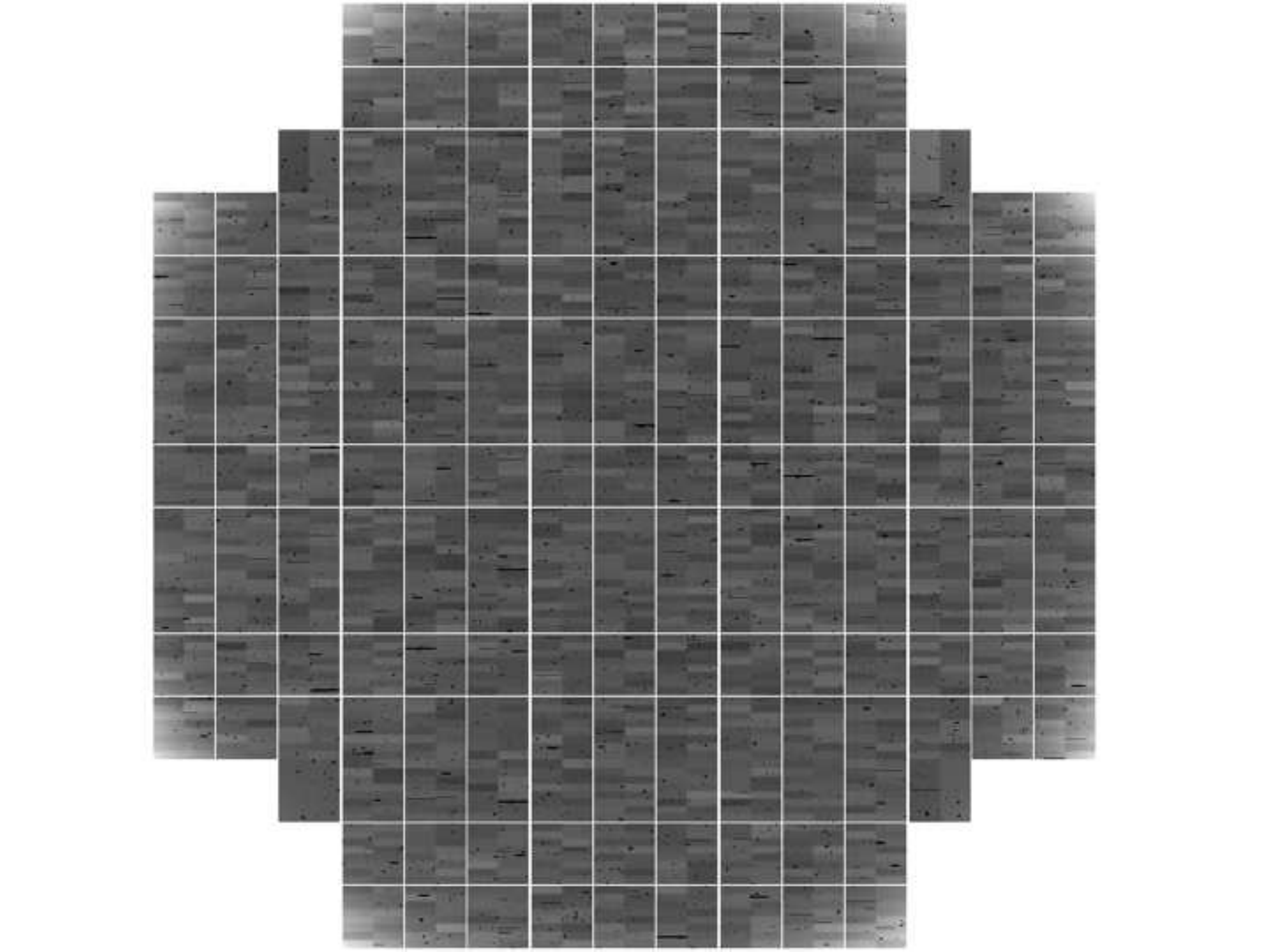}
\end{center}
\caption{\label{fig:label5}  A simulation of the entire field (10 sq. degrees)
of the LSST field of view.  10 million stars and galaxies are in the
simulation with over 1 trillion photons.  This simulation was executed using CONDOR grid
computing for about 1000 CPU hours in which each individual CCD was simulated
in parallel.  The image has over 3 billion pixel, so the full detail cannot be
observed.  The variation in bias levels of the individual amplifier is
visible, as well the vignetting of the background near the corners of the field.}
\end{figure*}

\begin{figure*}[htb]
\begin{center}
\includegraphics[width=0.9\columnwidth]{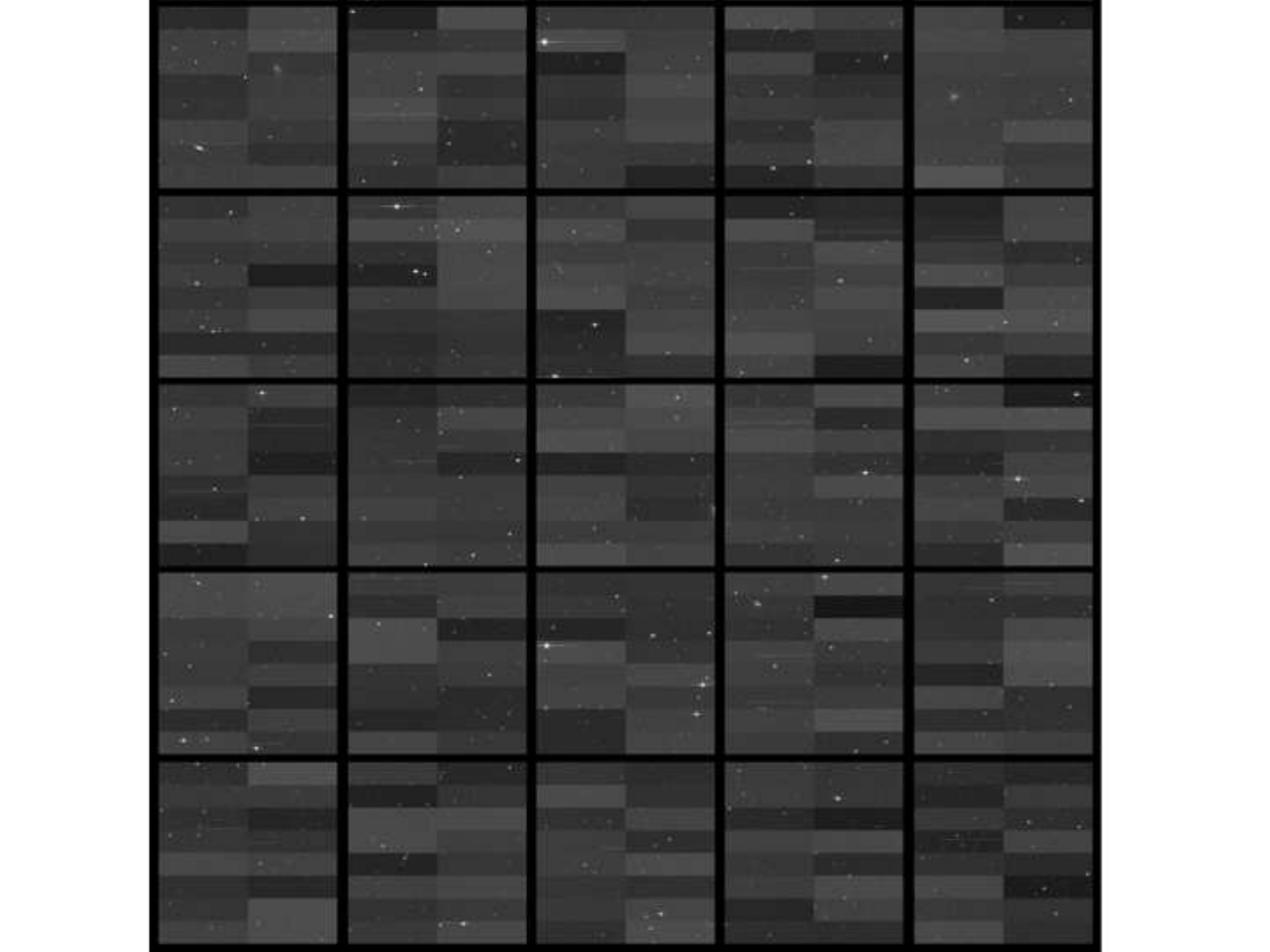}
\end{center}
\caption{\label{fig:label6}  The central 5x5 chips covering about 1
  sq. degree.  Individual bright stars are visible as well as hot columns.}
\end{figure*}

\begin{figure*}[htb]
\begin{center}
\includegraphics[width=0.9\columnwidth]{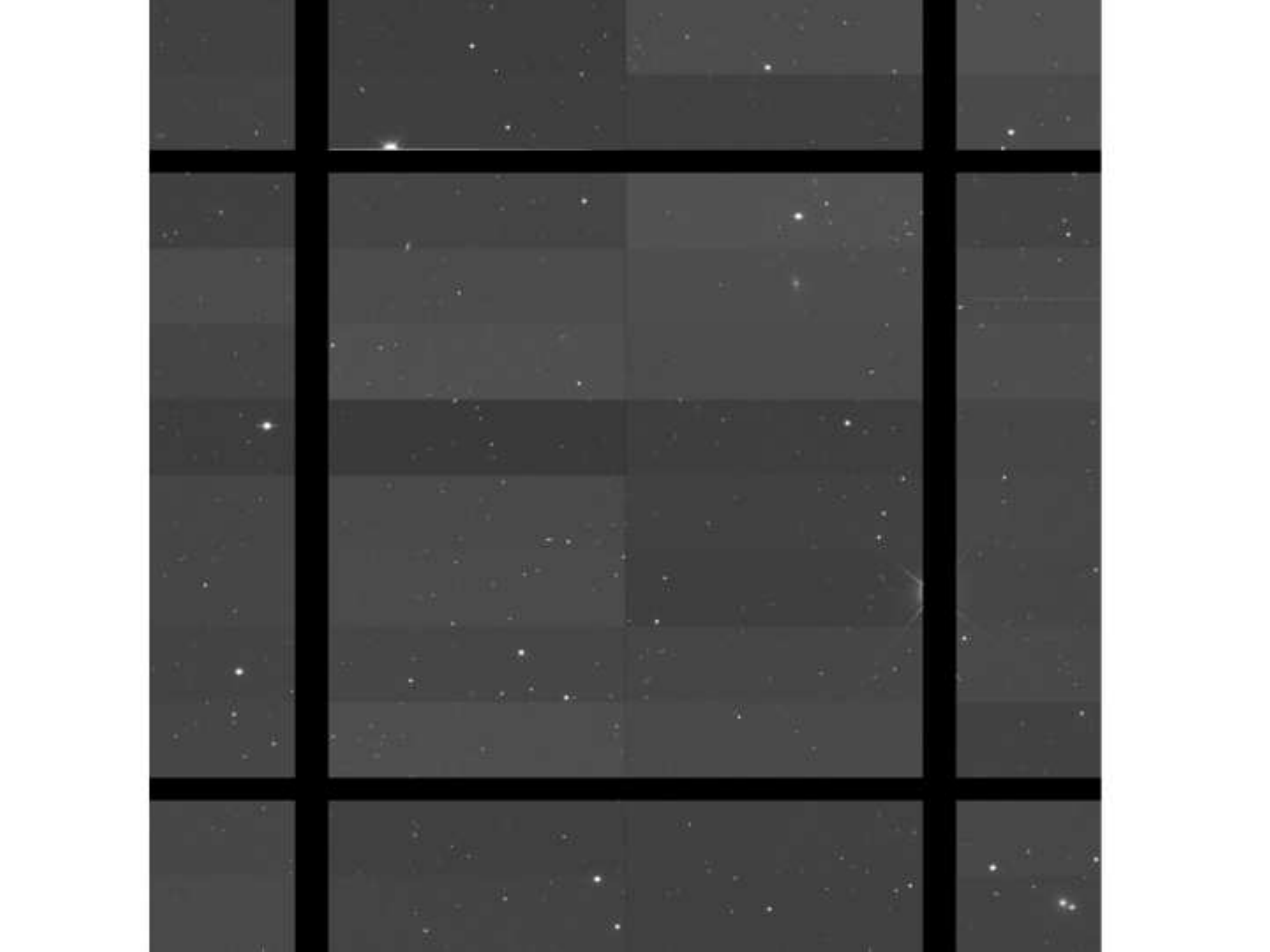}
\end{center}
\caption{\label{fig:label7}  A close up view of the central chip where more
  stars and brighter galaxies are visible.}
\end{figure*}

\begin{figure*}[htb]
\begin{center}
\includegraphics[width=0.9\columnwidth]{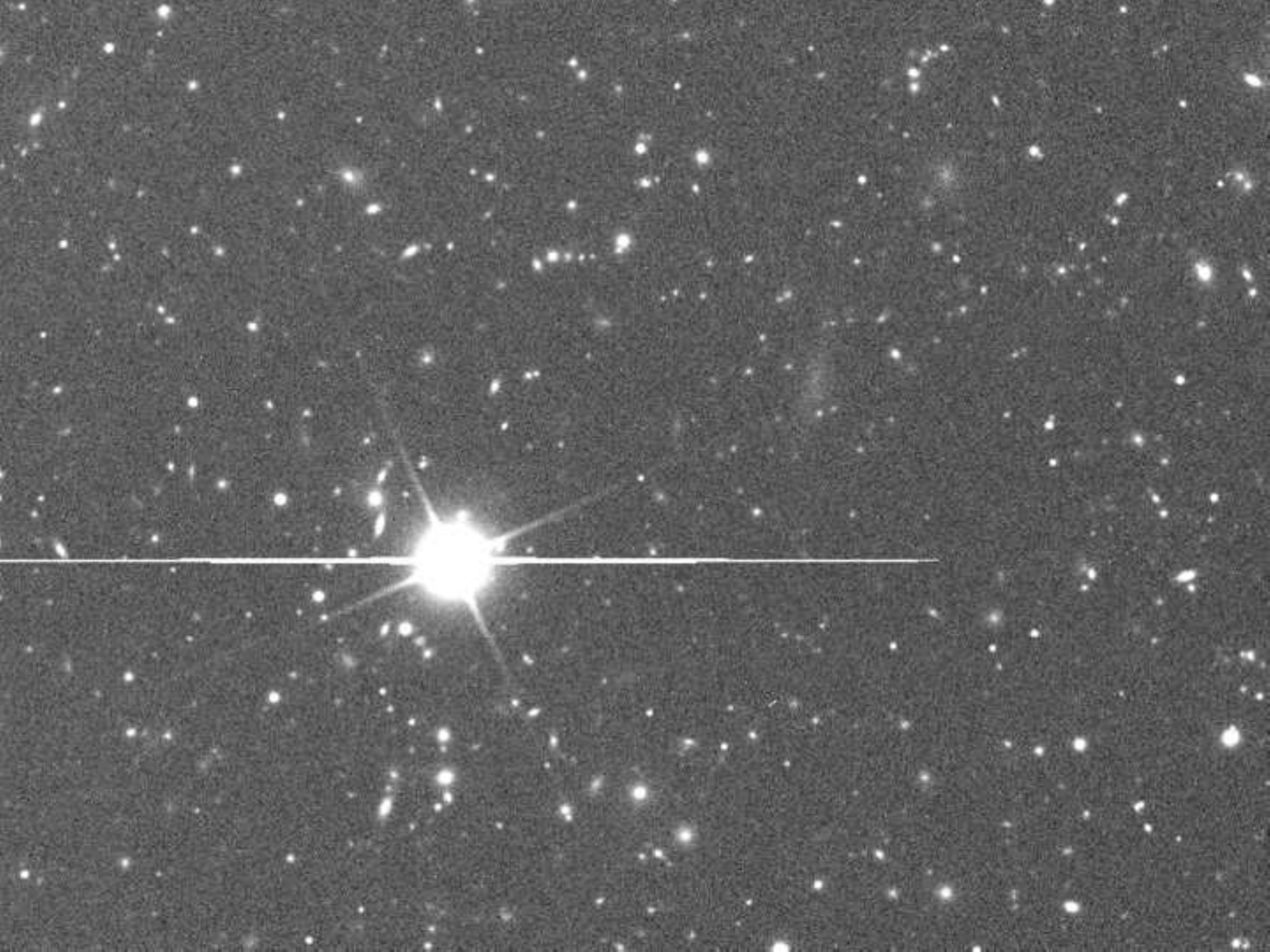}
\end{center}
\caption{\label{fig:label8}  An image of 3 amplifiers (3/16th of a chip)
  showing the full view of the stars and galaxies that we have simulated.
  Every photon has had the full physical simulation described in this work.
  Typically, an average galaxy near the detection threshold (24th magnitude)
  only has a few thousand photons in a 15 second exposure.}
\end{figure*}

\begin{figure*}[htb]
\begin{center}
\includegraphics[width=0.9\columnwidth]{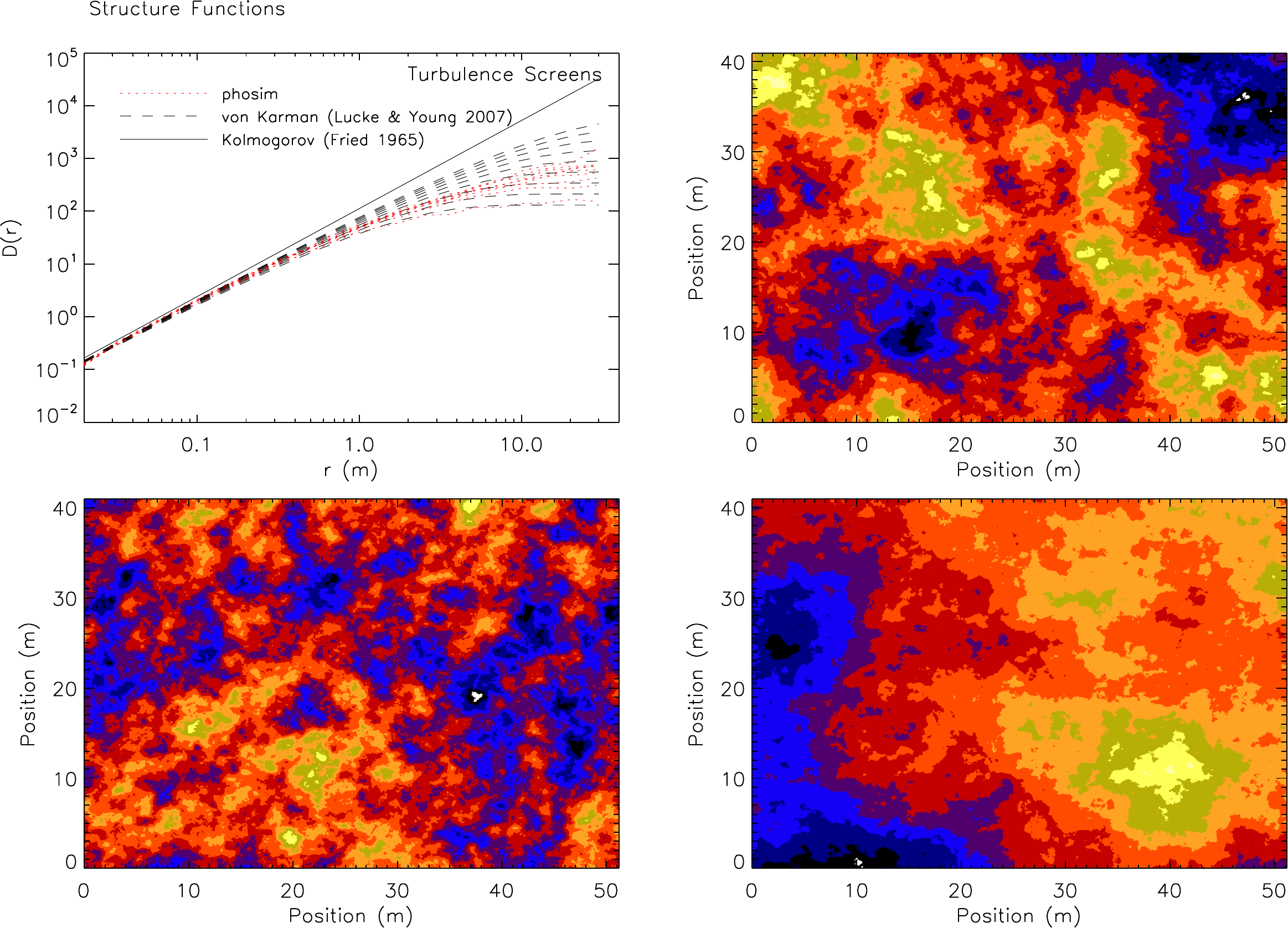}
\end{center}
\caption{\label{fig:label9}  Examples of three phase screens on 50m by 50m scales.  The colors indicate the relative phase shift as a function of position.  The images are generated by
adding the phase screens from the 7 atmosphere layers and each phase screen on the 4 different pixel scales for
each layer.  No obvious artifacts result from the numerical grids.  The upper left panel shows comparison
with the structure functions predicted for a pure Kolmogorov model (\citealt{fried1965}) and a von Karman model for different outer scales (\citealt{lucke2007})
}
\end{figure*}

\begin{figure*}[htb]
\begin{center}
\includegraphics[width=0.9\columnwidth]{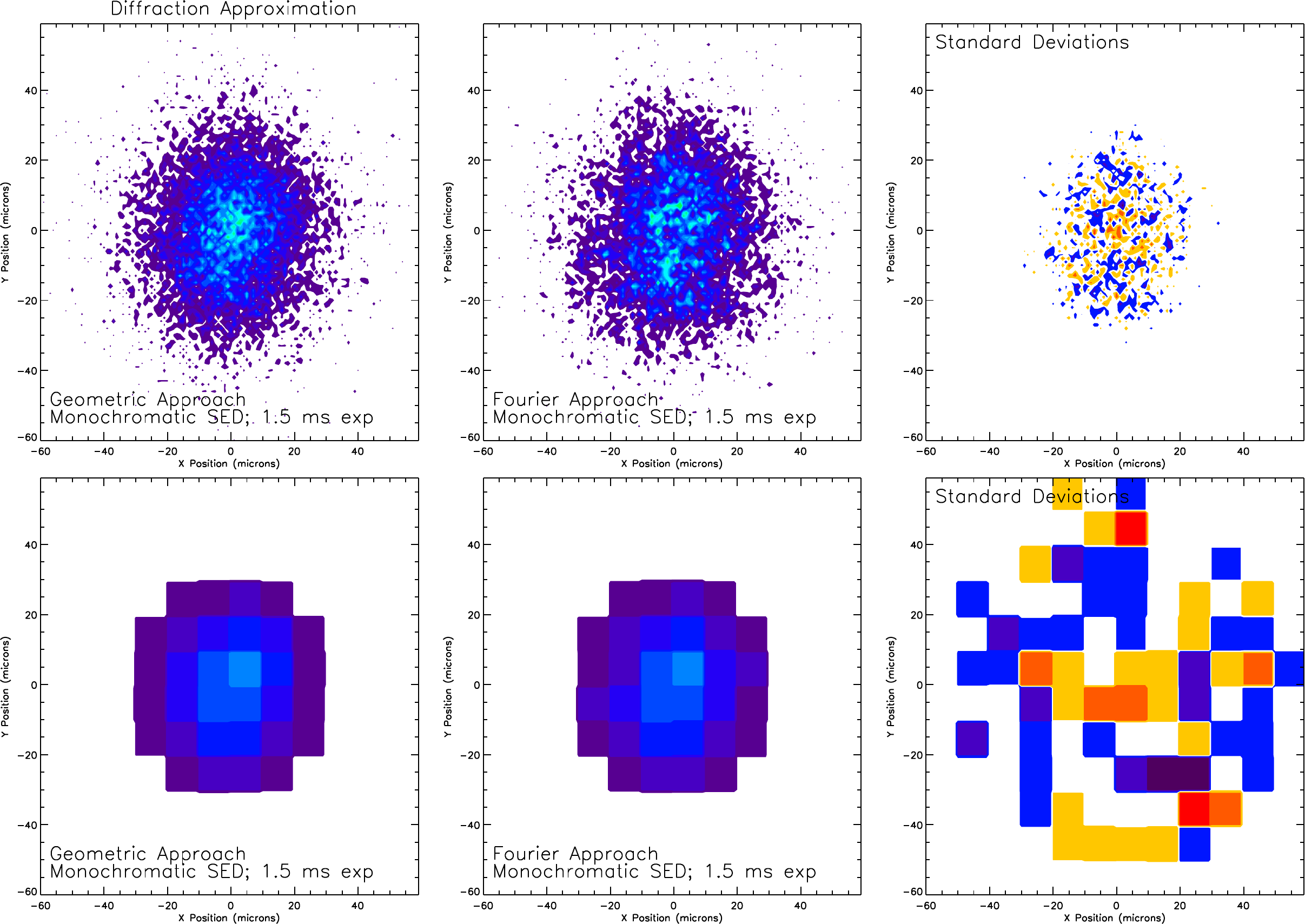}
\end{center}
\caption{\label{fig:label10}  A comparison of the point-spread-function induced
by the atmosphere using a full Fourier approach (middle panels) and the two
approximations discussed in the Appendix (left panels).  The simulations are
for a source exposed for 1.5 milliseconds.  The top panels show the PSF before
pixelization, and the right panels show the residual difference (colors
represent the -3 (dark purple), -2 (purple), -1 (blue), 1 (yellow), 2 (orange), 3 (red) sigma residuals).
  The only visible differences are the speckle structure that
occurs on 1/20th of a pixel scale.  The geometric approach PSF has FWHM of $0.800 \pm 0.021$,
ellipticity of  $-0.113 \pm 0.007$, $0.082 \pm 0.007$ and centroid of
$0.015 \pm 0.006$, $-0.003 \pm 0.006$.  The fourier approach PSF has FWHM of
 $0.834 \pm 0.021$, ellipticity of $-0.136 \pm 0.007$, $0.064 \pm 0.007$ and centroid of
$0.013 pm 0.006$, $-0.004 \pm 0.006$.  The reduced $\chi^2$ of the residual map is 1.36.}
\end{figure*}

\begin{figure*}[htb]
\begin{center}
\includegraphics[width=0.9\columnwidth]{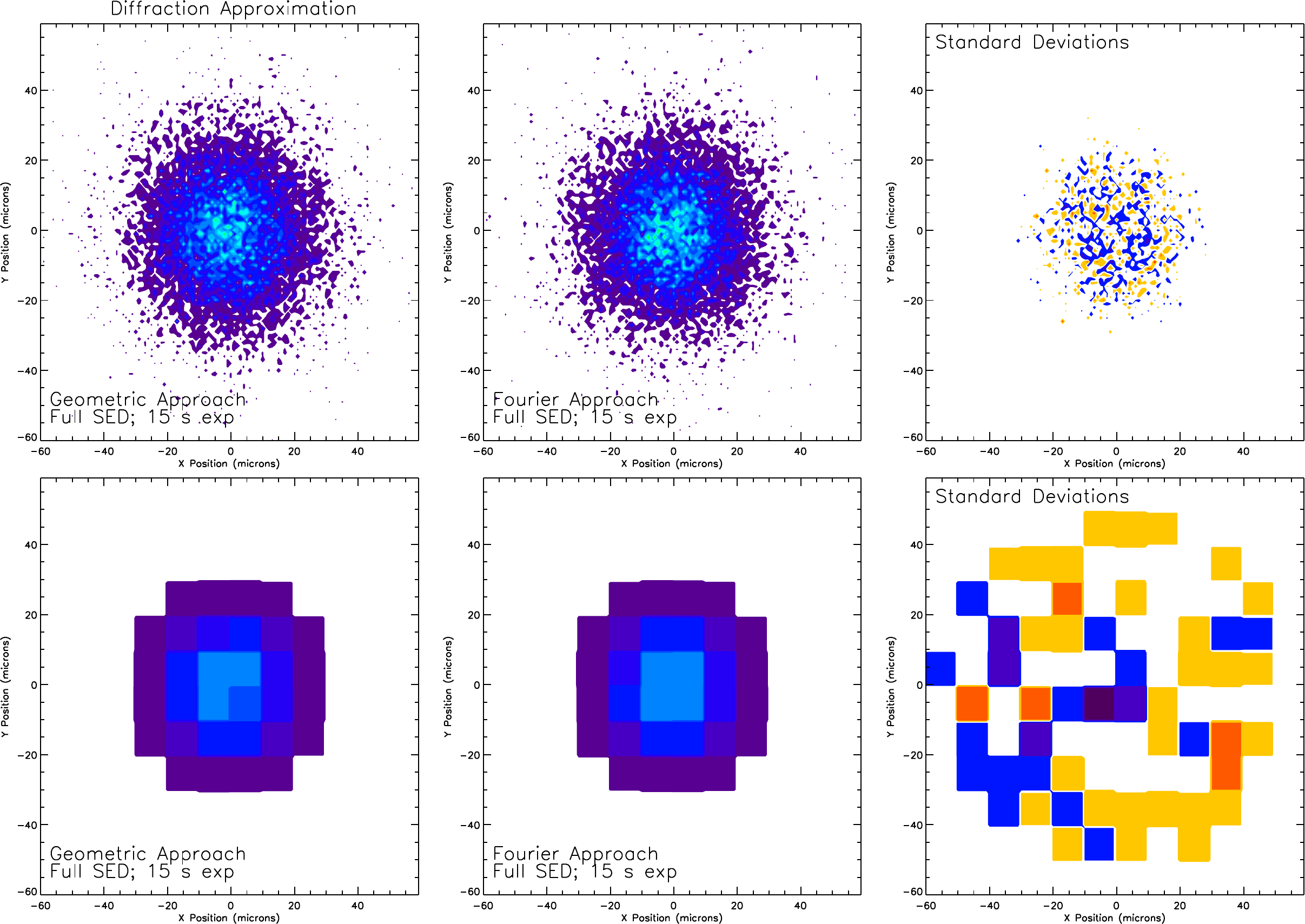}
\end{center}
\caption{\label{fig:label11}  The same as the previous figure except for a 15
  second exposure.  The difference in the two simulation methods is purely statistical.
The geometric approach PSF has FWHM of $0.0793 \pm 0.021$,
ellipticity of  $-0.015 \pm 0.007$, $-0.016 \pm 0.007$ and centroid of
$-0.007 \pm 0.006$, $0.003 \pm 0.006$.  The fourier approach PSF has FWHM of
 $0.766 \pm 0.021$, ellipticity of $-0.011 \pm 0.007$, $0.005 \pm 0.007$ and centroid of
$-0.009 pm 0.006$, $-0.001 \pm 0.006$.  The reduced $\chi^2$ of the residual map is 0.84.}
\end{figure*}

\begin{figure*}[htb]
\begin{center}
\includegraphics[width=0.9\columnwidth]{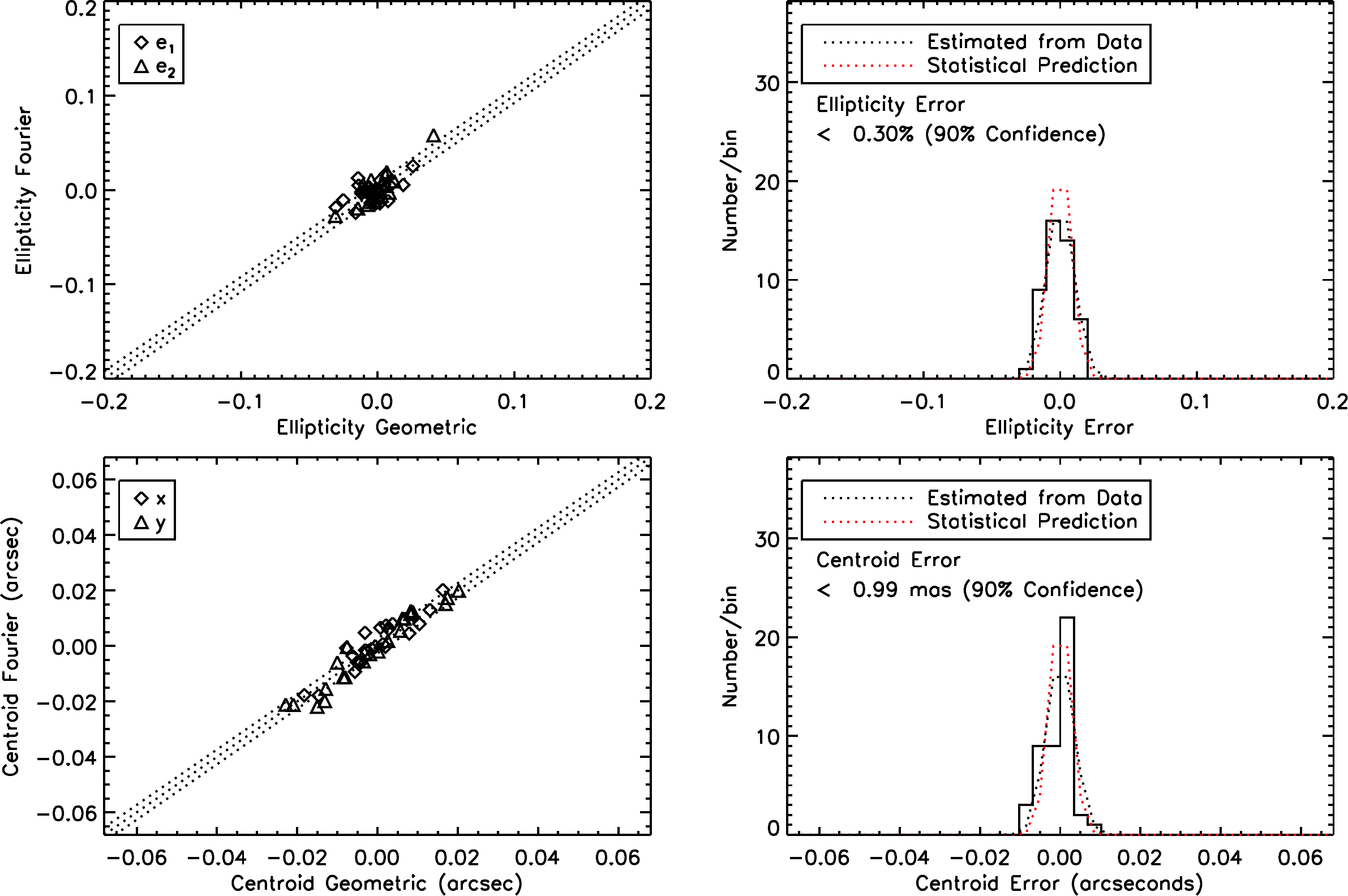}
\end{center}
\caption{\label{fig:label12}  Measurements of the centroid and ellipticity of example atmospheres using the two
techniques.  The y-axis of the left plot uses the traditional Fourier light propagation technique, whereas the
x-axis is the measurement using the geometric hybrid approach.  There is a clear correlation demonstrating that
both techniques are capturing the same details.  The statistical differences are shown in the right panels.  The distribution is consistent with expected statistical errors (red), and can be used to set an upper limit to the
numerical technique (see text).  }
\end{figure*}

\begin{figure*}[htb]
\begin{center}
\includegraphics[width=0.9\columnwidth]{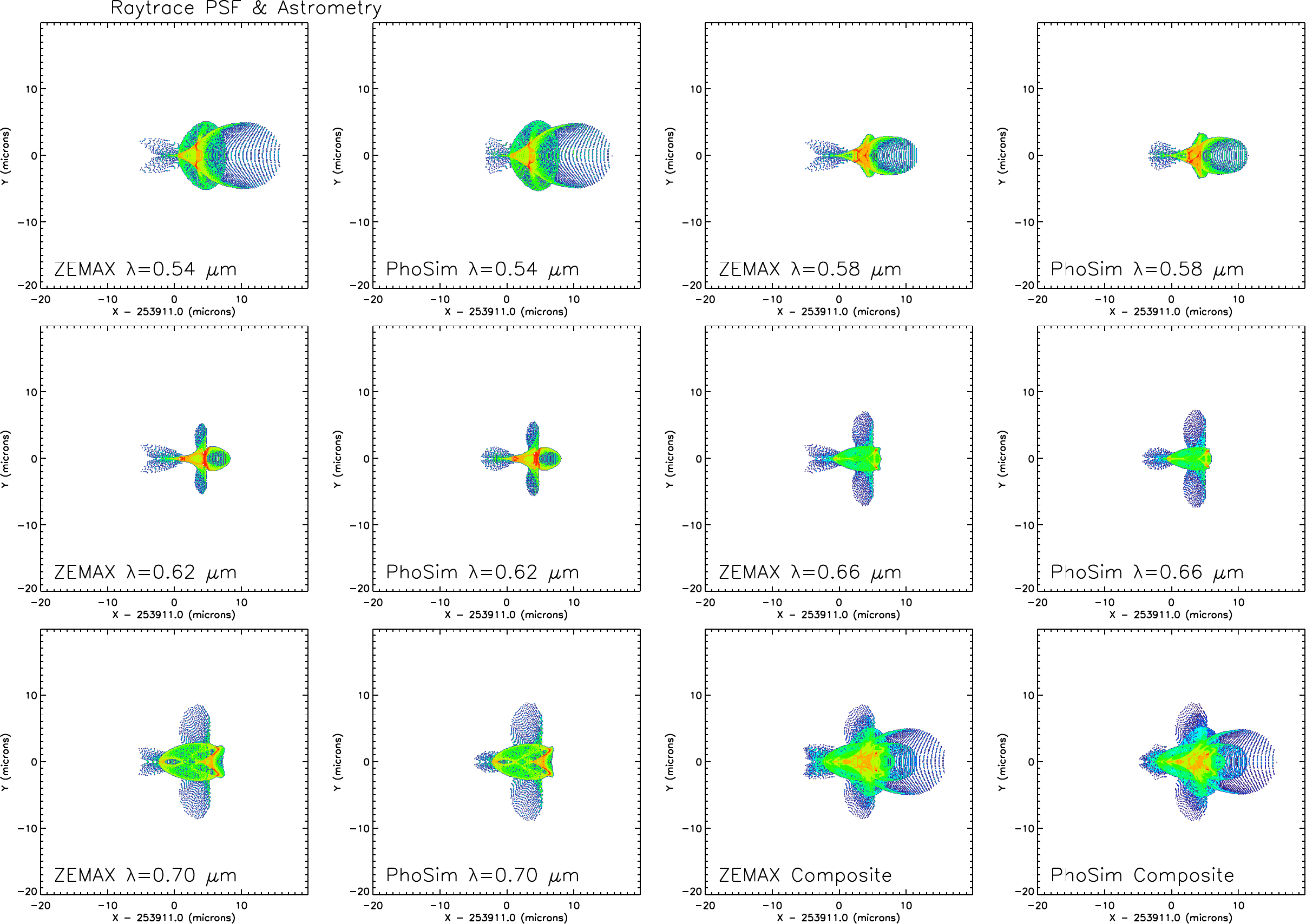}
\end{center}
\caption{\label{fig:label13}  Comparison of spot diagrams through the LSST
  design for an off-axis star at various wavelengths.  Note the scale makes
  these images fill only 4x4 LSST pixels, so the detail is reproduced on a
  very fine scale.  The x-axis has a large offset subtracted so this demonstrates that the rays are arriving at the focal plane to a very high accuracy.  Detailed ray by ray comparison shows a rms difference of 0.018 microns, which is negligible for all practical science cases.}
\end{figure*}

\begin{figure*}[htb]
\begin{center}
\includegraphics[width=0.9\columnwidth]{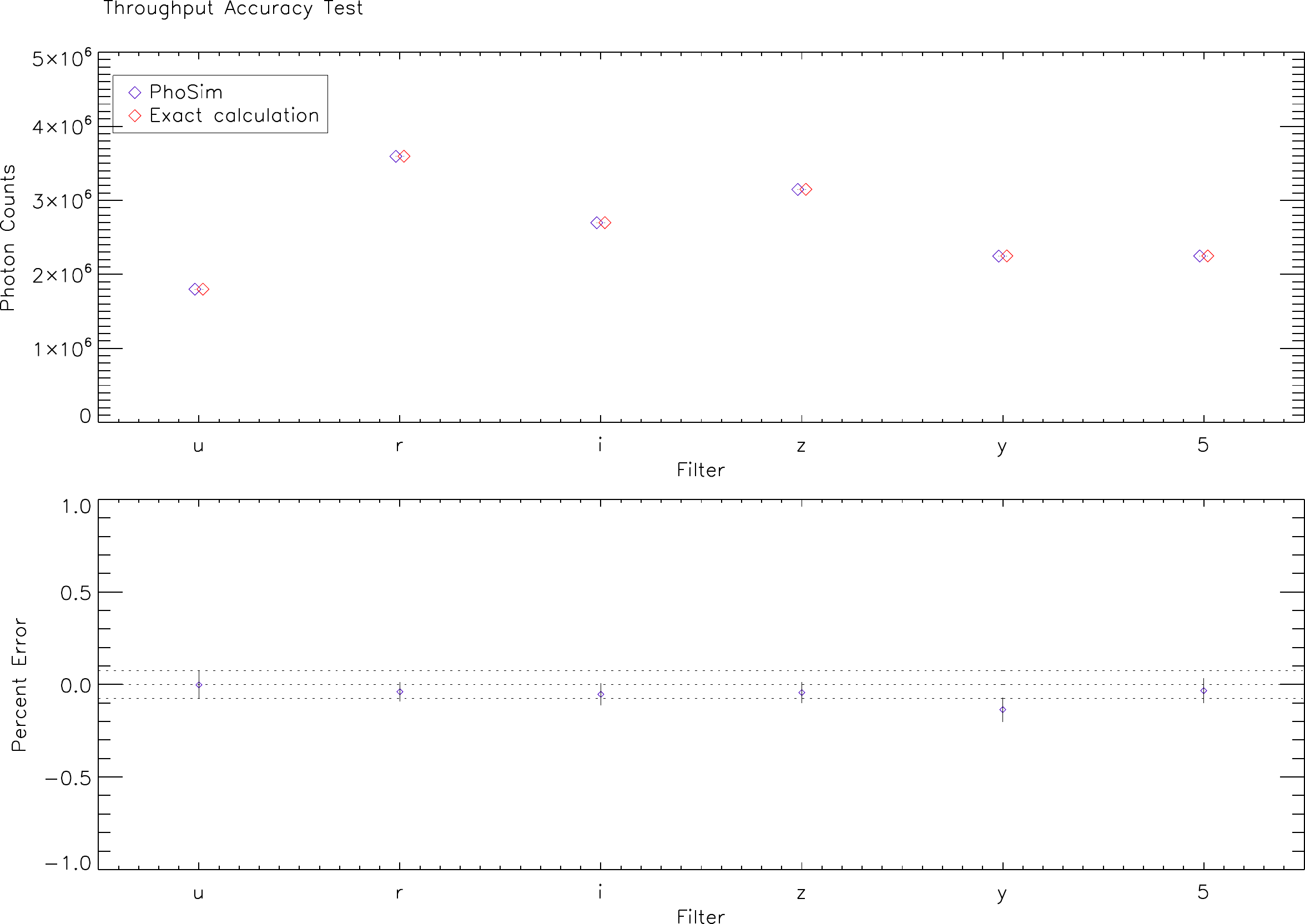}
\end{center}
\caption{\label{fig:label14}  The photon throughput using six different square spectral energy distributions in PhoSim vs. an exact analytic photometry integral (see text).  The bottom plot shows the residual photometric error, and tests PhoSim's ability to properly sample SEDs, accept or reject photons through various modules,
simulate the correct geometric acceptance.}
\end{figure*}

\begin{figure*}[htb]
\begin{center}
\includegraphics[width=0.9\columnwidth]{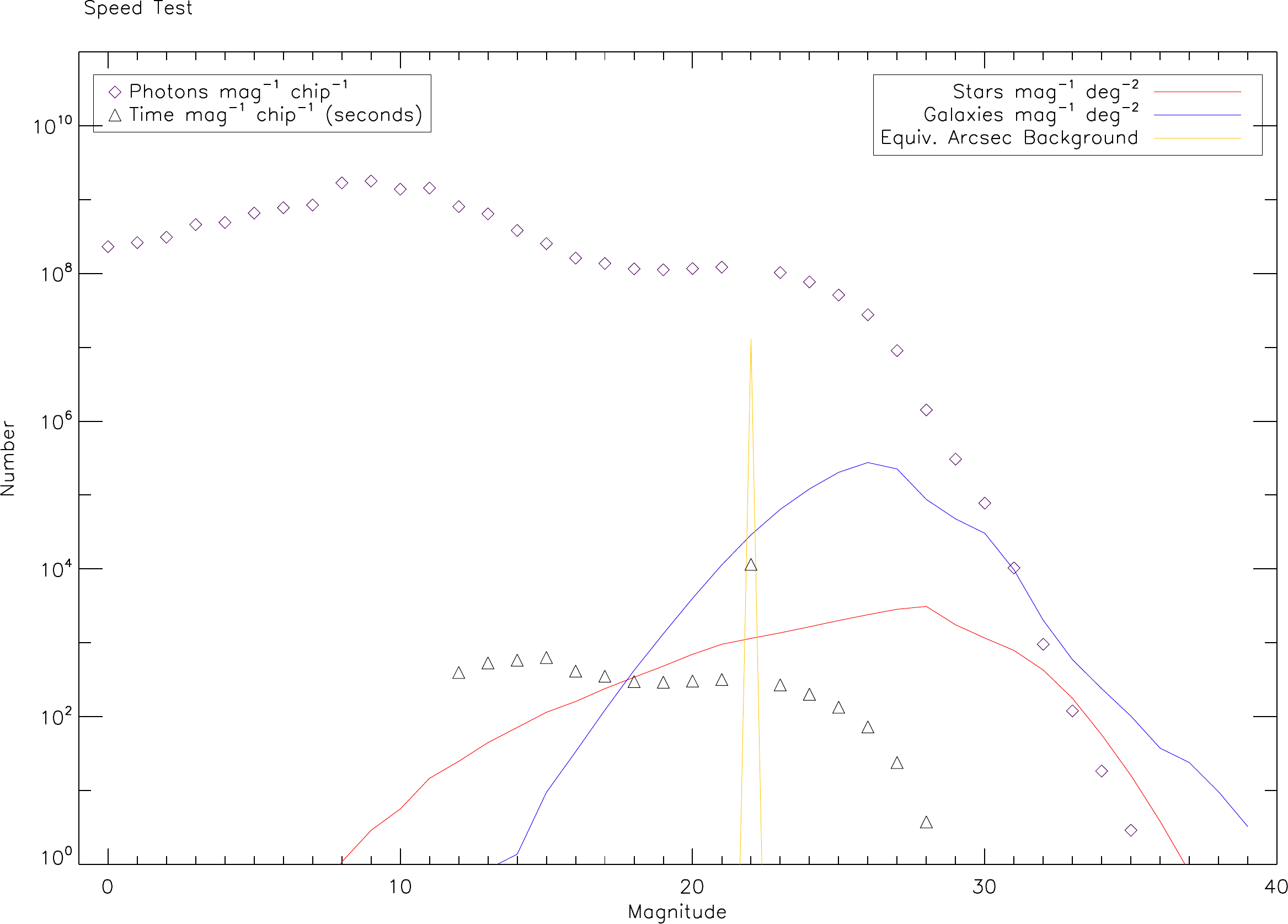}
\end{center}
\caption{\label{fig:label15}  The speed of the simulator.  The diamonds show
  the number of photons from all astrophysical sources on a LSST chip (13 by
  13 arcminutes).  The blue and red curves show how many stars and galaxies,
  respectively, per sq. degree.  The yellow curve splits the sky background
  into equivalent sources placed every 15 arcseconds.  The number of seconds for
  each magnitude bin is shown by the triangles.  For faint non-background
  unsaturated sources the simulation time scales as the number of photons and is typically 400,000 photons per second.
  For saturated sources and background there are significant optimizations
  that increase the simulation time.  A 12th magnitude source is simulated at about a factor of 5 faster than an unsaturated source, and the background is simulated about a factor of 50 faster than unsaturated sources.
  There is typically less than 1 source per LSST chip (13 by 13 arcminutes) below 12th magnitude, so those sources are not common.}
\end{figure*}

\end{document}